\documentclass[%
 reprint,
superscriptaddress,
 amsmath,amssymb,
aps,
prb,
citeautoscript,
]{revtex4-1}

\usepackage{graphicx}
\usepackage{dcolumn}
\usepackage{bm}
\usepackage{hyperref}
\hypersetup{
    colorlinks = true,
    allcolors=black,
    citecolor=blue,
    linkcolor=blue
} 
\usepackage{amsmath, amsfonts, dsfont, mathrsfs, float,amssymb}
\usepackage{color}
\usepackage{cancel}
\usepackage{tensor}
\usepackage{setspace}
\usepackage{ragged2e}
\usepackage{verbatimbox}
\usepackage{tikz}
\usepackage{multirow,tabularx}
\usetikzlibrary{calc}
\usetikzlibrary{shapes.geometric}
\usetikzlibrary{arrows}

\newcommand{\dd}{\text{d}}

\newcommand{\barj}[1]{\bar{\jmath}}

\renewcommand{\exp}[1]{e^{#1}}

\newcommand{\rfig}[1]{Fig. \ref{fig:#1}}

\begin{document}


\title{Information-Theoretic Memory Scaling in the Many-Body Localization Transition}

\author{Alexander Nico-Katz}
\affiliation{Department of Physics and Astronomy, University College London, London WC1E 6BT, United Kingdom}
\author{Abolfazl Bayat}
\affiliation{Institute of Fundamental and Frontier Sciences, University of Electronic Science and Technology of China, Chengdu 610051, China}
\author{Sougato Bose}
\affiliation{Department of Physics and Astronomy, University College London, London WC1E 6BT, United Kingdom}

\date{\today}
             
\begin{abstract}
    A key feature of many-body localization is the breaking of ergodicity and consequently the emergence of local memory; revealed as the local preservation of information over time. As memory is necessarily a time dependent concept, it has been partially captured by a few extant studies of dynamical quantities. However, these quantities suffer from a variety of issues which limit their value as true quantifiers of memory; and thus a fundamental and complete information-theoretic understanding of local memory in the context of many-body localization remains elusive. We outline these issues in detail and introduce the dynamical Holevo quantity to address them. We find that it shows clear scaling behavior across the many-body localization transition, and we determine a family of two-parameter scaling ans\"atze which capture this behavior. We perform a comprehensive finite size scaling analysis to extract the transition point and scaling exponents.
\end{abstract}
             
\maketitle


\section{Introduction}
\label{sec:intro}

How many bits of information stored locally in a quantum many-body system are preserved over time? The most striking scenario in which to ask this question is in the context of many-body localization (MBL). In MBL systems, quenched disorder frustrates natural, scrambling, self-thermalizing dynamics \cite{Nandkishore2015, Abanin2019, Alet2018} leading to the local preservation of information: the emergence of memory. Unlike conventional quantum phase transitions~\cite{sachdev2011}, the MBL transition takes place across the spectrum~\cite{Abanin2017, Abanin2019, Nandkishore2015, Alet2018}; making its analysis a far more elaborate task than that of other quantum critical systems. Despite this, several features of the MBL phase have been characterised, including Poisson-like level statistics \cite{Oganesyan2007,Pal2010,Kjall2018,Sierant2017,Roushan2017, Song1999}, area-law entangled eigenstates~\cite{Eisert2010,Grover2014,Bauer2013,Kjall2014,Khemani2017,Khemani2017a}, slow growth of correlations with time \cite{Kjall2018,Lukin2019,Wei2018,BarLev2015}, and the breakdown of transport~\cite{Iemini2016,Lezama2017,Schreiber2015,Choi2016,Rubio-Abadal2019,Sierant2017,Xu2018,Bordia2017}. To identify these features, various quantities have been exploited, including quantum mutual information \cite{DeTomasi2017}, Schmidt gap \cite{Gray2018}, entanglement in the form of concurrence~\cite{Wei2018,Bera2016, Iemini2016}, entropies~\cite{Khemani2017,Khemani2017a,Lim2016,Kjall2014,Luitz2015,Zhang2018,Vosk2015,Kjall2018,Roushan2017,Bera2015,Iemini2016,Ponte2015,Schreiber2015,Bardarson2012,Znidaric2008,Serbyn2015,Wei2018,Xu2018,Pietracaprina2017} and negativity \cite{Gray2018, Gray2019, Wei2018}, population imbalance~\cite{Iemini2016,Lezama2017,Schreiber2015,Choi2016,Rubio-Abadal2019,Sierant2017,Xu2018,Bordia2017,Luitz2016} and other occupancy-like quantities \cite{Pal2010, Lim2016,Ponte2015,Lezama2017,Luitz2015,Bera2015,Smith2016,Serbyn2015}.

The above works are either concerned with spatial correlations or are missing a bitwise interpretation, and do not fully capture the temporal preservation of information; i.e. memory. Extant studies of dynamical quantities, primarily entanglement growth and the population imbalance~\cite{Pal2010, Luitz2016, Ponte2015, Bordia2017, Smith2016}, only  partially capture memory. For example, the imbalance and similar quantities are dependent on the measurement basis, and a sub-optimal choice can obscure otherwise accessible information.
\begin{figure}[ht]
        \includegraphics[width=\linewidth]{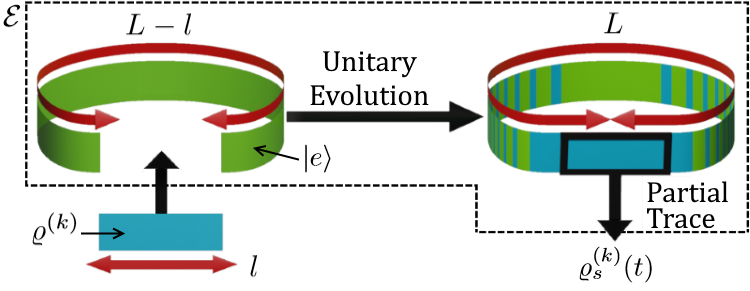}
        \caption{Schematic diagram of the procedure by which individual messages $\varrho^{(k)}$ are transmitted via the map $\mathcal{E}$. Information initially localized within the message may bleed out into the environment during transmission.}
        \label{fig:fig0}
\end{figure}
Raw informational quantities like entanglement entropy or spatial correlation functions account for this, but fail to distinguish between input states or quantify the amount of \textit{accessible} information in a block. The entanglement entropy in particular also lacks a temporal component: simply giving us an indication of the instantaneous mixedness of a subsystem. Thus, it is highly desirable to have a complete, unbiased, information-theoretic quantification of local memory in the context of MBL.


The  investigation of the MBL phase and the many-body localization transition (MBLT) is broadly conducted through two different classes of quantities: (i) static quantities computed over many-body eigenstates (often selected from a small energy interval)~\cite{Gray2018,Oganesyan2007,Pal2010,Kjall2014,Bera2016,Kjall2018,Khemani2017,Lim2016,Khemani2017a,Luitz2015,Zhang2018,Vosk2015,Song1999,Sierant2017,Roushan2017,Bera2015,Serbyn2015,DeTomasi2017,Gray2019,Sierant2020pol}; and (ii) dynamical quantities computed over the time-evolved quantum state of a system which has overlap with several eigenstates~\cite{Kjall2014,Iemini2016,Ponte2015,Schreiber2015,Bardarson2012,Znidaric2008,Serbyn2015,Yao2016,Wei2018,Xu2018,Rubio-Abadal2019,Choi2016,Lezama2017,Sierant2017,Bordia2017,Lukin2019,BarLev2015,Pal2010,Smith2016,Luitz2016,Bera2017,Doggen2018,Chanda2020,Chanda2020b}. Scaling near the MBLT has been 
investigated mainly through static quantities such as the level statistics~\cite{Song1999,Luitz2015,Sierant2017} and  entanglement entropy~\cite{Kjall2014,Khemani2017a,Luitz2015,Zhang2018,Vosk2015}. Investigating the properties of scaling through dynamical quantities is more challenging and less thoroughly explored (see e.g. Ref.~\cite{Kjall2014,Potter2015,Lezama2017}). Memory is necessarily a dynamical quantity, motivating three main questions: (i) what is required of a quantifier for it to be a quantifier of local memory? (ii) can one construct such a quantifier which is optimal in the sense that it is independent of measurement basis and captures the maximum possible amount of accessible information? And (iii) if so, what it its behaviour across the MBLT, and does it exhibit scaling? Addressing these questions is crucial to developing an informational understanding of the nature of the MBLT.


In this paper, we address these questions by discussing local memory and introducing criteria which a quantity must satisfy to be a true memory quantifier. We introduce a dynamical version of the Holevo quantity as a complete and optimal information-theoretic memory quantifier. We investigate it across the MBLT and perform a comprehensive scaling analysis over its late-time values using a family of two-parameter scaling ans\"atze, from which we extract critical parameters.


\section{Memory in Many-Body Localized Systems}

The notion of `local memory' is widely quoted in the literature of MBL, but is infrequently the subject of direct investigation. In this section we outline and discuss the ways in which memory has been captured in MBL systems before, and leverage this discussion into a determination of the important features that a quantifier of memory must have - features that none of the extant quantities display completely.

In theoretical and experimental studies alike, memory is most frequently discussed in terms of non-zero steady-states of appropriate observables; notably quantities derived from local magnetization or occupancy measurements \cite{Pal2010, Nandkishore2015, Ponte2015, Smith2016, Chandran2015liom, VanHorssen2015, Brenes2018, Kuno2020}. Should these measurements systematically coincide with similar measurements made on the initial state of the system, then system has retained some `local memory' of those initial features. A generic quantity of this kind is the autocorrelation function:
\begin{equation}
    F(t) = \left\langle W(0)W(t) \right\rangle
\end{equation}
of some appropriate observable $W(t) = \langle\hat{W}(t)\rangle$.

The premier example of this type of quantity is the imbalance, used extensively in MBL literature and to great effect in landmark experiments (see, for example, Ref.~\cite{Schreiber2015, Luschen2017, Bordia2017, Xu2018}). It is usually defined in terms of local fermionic number expectation values $\langle\hat{n}_j(t)\rangle = n_j(t)$ where $j$ indexes sites on a lattice. If the initial system is in some charge density wave configuration then the aggregate deviation of the $n_j(t)$ from their initial values $n_j(0)$ quantifies how well the system remembers its initial number configuration. The prototypical example, for a system of spinless fermions such that $n_j(t) \in [0, 1]$ and initialised in the Fock state $|0,1,0,\cdots,0,1\rangle$ (a charge density wave configuration), the imbalance is defined as:
\begin{equation}
    \mathcal{I}(t) = \frac{N_e(t) - N_o(t)}{N_e(t) + N_o(t)}
\end{equation}
where $N_{e(o)} = \sum_{j \in \text{even}(odd)}n_j(t)$ is the total number of fermions on even (odd) sites. The initial state has all even sites unoccupied, and all odd sites occupied, so $\mathcal{I}(0) = 1$. As the system evolves, it can either thermalize to homogeneity such that the initial configuration is lost $\lim_{t \to \infty} \mathcal{I}(t) = 0$, else it can relax to be correlated $\lim_{t \to \infty} \mathcal{I}(t) > 0$ or anti-correlated $\lim_{t \to \infty} \mathcal{I}(t) < 0$ with its initial state. The prevailing issues with the use of such quantities as memory quantifiers are twofold: first they have no bitwise interpretation, and second a poor choice of measurement basis can obscure otherwise accessible information. To illustrate the latter case consider a protocol which perfectly transmits $Z$-basis eigenstates to $X$-basis eigenstates: $|0\rangle \to |+\rangle$, $|1\rangle \to |-\rangle$, but where the final measurements on these states are made in the $Z$-basis. This issue means that even if informational versions of local observables are constructed, they still do not give a complete understanding of how much information has been retained.

A more sophisticated grasp of memory from the perspective of information scrambling can be attained by investigating the growth of the out-of-time-order correlator (OTOC)
\begin{equation}
    O^{WV}(t) = \left\langle [W(t), V(0)]^\dagger [W(t), V(0)] \right\rangle_\beta
\end{equation}
for some appropriately chosen, spatially distant, operators $\hat{W}$ and $\hat{V}$, which originally commute. The $\langle\cdot\rangle_\beta$ here denotes the thermal average at inverse temperature $\beta$. Originally envisaged as an analogy to the classical Poisson bracket as a measure of quantum chaos, it can also be interpreted as an indirect measure of information scrambling: the speed and strength with which the effect of the perturbation $\hat{V}$ is felt by the distant $\hat{W}$ tells us how quickly information is carried through the system. In the ergodic phase, the effect of the perturbation spreads rapidly and the OTOC grows exponentially in time $O^{WV}(t) \sim e^{\lambda_L t}$ at a rate governed by the Lyapunov exponent $\lambda_L$; whilst in the localized phase this growth appears logarithmic or power-law \cite{Chen2016,Lee2019,Xu2019}. The OTOC approach, whilst more nuanced, depends on an appropriate choice of operators, has no clear interpretation in terms of how much information can actually be extracted from a given subsystem, and is exceedingly difficult to measure experimentally.

Finally, local memory can be inferred without appealing to time correlation by monitoring, e.g., the growth of entanglement entropies, the spatial mutual information, and the extraction of local integrals of motion \cite{DeTomasi2017, Chandran2016, Yu2016, Pekker2017, Villalonga2020}. Some of these quantities have obvious bit-wise informational interpretations, or are advantageously blind to the specifics of measurement procedure. Despite this, all have shortcomings which curtail their use as true memory quantifiers. Entanglement entropies quantify the instantaneous mixedness of a  subsystem, and thus capture how valuable they are as instantaneous alphabets, but not how much accessible information is actually stored in them with respect to an initial message. The same issue exists in the context of extraction of local integrals of motion and their physical extent which, though exceedingly valuable as direct probes of the MBL regime itself, are difficult to probe experimentally, contain no clear correlation to initial information distributions, and lack bitwise interpretations. The spatial mutual information also suffers from this lack of temporal correlation: it quantifies how separate subsystems correlate, but not how well they correlate with their own past.

In summary, the prevailing methods by which memory is accessed in MBL systems all have respective strengths and shortcomings. The dynamics of local observables like the magnetization and imbalance are experimentally tractable and temporally connect the initial conditions with late-time measurements; but can be rendered useless by a poor choice of measurement basis, and do not quantify how much information - in bits - can be extracted from a subsystem. OTOCs, whilst much more sophisticated and theoretically invaluable, suffer similarly from the specification of perturbation/measurement operators and the lack of a bitwise interpretation, and are not readily accessible to experiment. Quantities like the entropy and spatial mutual information are informational, but lack the temporal correlations necessary to act as a true memory quantifier. 

Based on this analysis, we define two requirements for a quantity to be considered a true quantifier of memory. A memory quantifier must have: (i) temporal correlations which relate the initial and final states of an appropriately defined message register, and (ii) a bit-wise interpretation of the amount of information a subsystem has retained. We also state two preferred features which, whilst not necessary, are advantageous to a quantifier: (i) optimal, in the sense that no change in measurement basis increases the amount of information captured, and (ii) experimental accessibility.

\section{The Holevo Quantity} 
We introduce the Holevo quantity to address the requirements outlined in the previous section. The Holevo quantity quantifies the amount of classical information, in bits, which can be accessed via optimal measurements on an ensemble of information bearing quantum states \cite{holevo1973, nielsen2011, Roga2010}. For an ensemble of $M$ input states $\{ \varrho^{(1)},\varrho^{(2)},\cdots,\varrho^{(M)} \}$ undergoing a general quantum evolution in the form of a trace preserving completely positive map $\mathcal{E}$, the Holevo quantity is defined as
\begin{equation}\label{eq:Holevo_def}
C(t) = S\left(\sum_k p_k \mathcal{E}\left[\varrho^{(k)}\right]\right) - \sum_k p_k S\left(\mathcal{E}\left[\varrho^{(k)}\right]\right)    
\end{equation}
where $p_k$ is the probability with which the input $\varrho^{(k)}$ is sent through the map $\mathcal{E}$ and $S(\cdot)=-\text{Tr} \left[\cdot\log_2 \cdot\right]$ is the von Neumann entropy. This quantity is widely used for bounding the capacity of classical communication across a distance using quantum carriers \cite{holevo1973,Schumacher1997,Giovannetti2005,Macchiavello2004,Lupo2011,Yang2011,Banchi2017,D'Arrigo2015}.  Here we use this as a quantifier of memory, which can be regarded as a `communication in time'. The Holevo quantity has two distinct informational advantages: (i) it is optimal with respect to measurement basis \cite{Roga2010}; and (ii) it distinguishes between different input states $\varrho^{(k)}$ by construction. The temporal correlation which we posit as a necessary condition for a quantity to be a memory quantifier is between the initial ensemble (encoded in the $p_k$) and final ensemble (encoded by the $\rho^{(k)}$). The Holevo quantity is clearly informational, explicitly yielding the number of classical bits which remain accessible over time. Finally we note that the sums in Eq.~\ref{eq:Holevo_def} run over an ensemble of message states rather than individual sites, making it qualitatively different from conventional bulk correlation functions. These features together satisfy both the necessary conditions for a memory quantifier introduced at the end of the previous section. This makes the Holevo quantity a more viable and complete quantifier of memory than quantities like the imbalance and entanglement entropy, and one which is more experimentally accessible and informationally complete than the use of OTOCs, correlation functions, or explicit extraction of LIOMs.
\begin{figure}[ht]
        \includegraphics[width=\linewidth]{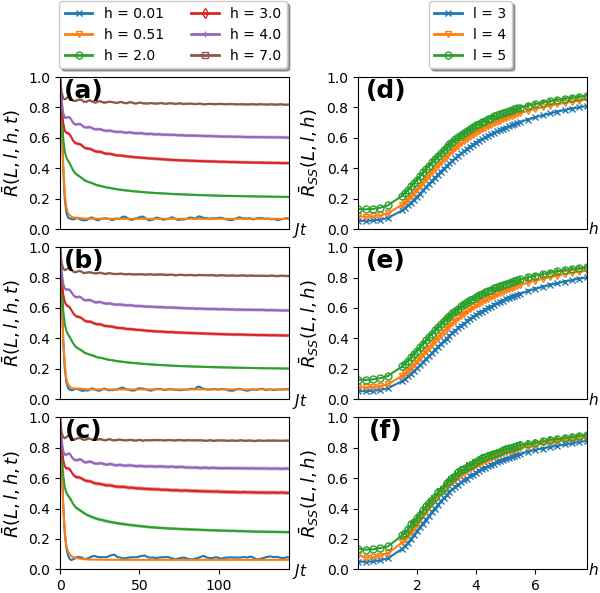}
        \caption{\textbf{(a,b,c)} The disorder-averaged Holevo rate $\bar{R}(L,l,h,t)$ against time for a fixed message length $l=4$. \textbf{(d,e,f)} the time-averaged Holevo rate $\bar{R}_{SS}(L,l,h)$ against disorder strength for a variety of message lengths. Each row contains results for a single environment type: \textbf{(a,d)} Neel state, \textbf{(b,e)} evolved Neel state, and \textbf{(c,f)} eigenstate environments, respectively. We take $L=16$ for all above figures.}
        \label{fig:fig1}
\end{figure}
\section{Model} 
We consider a system of $l$ spin-1/2 particles which encode pure separable messages of the form $\varrho^{(k)}=|m_1^{(k)},m_2^{(k)},\cdots,m_l^{(k)}\rangle \langle m_1^{(k)},m_2^{(k)},\cdots,m_l^{(k)} |$ in which $
m_i^{(k)}=0,1$ represents spin up and down respectively. This system is embedded in an environment of size $L-l$ which is initially prepared in a pure quantum state $|e\rangle$. The combined state of message and environment is of size $L$, and is initially given by the quantum state $\varrho^{(k)}_{se}(0)=\varrho^{(k)} \otimes |e\rangle \langle e|$.
The interactions between the particles are explained by the Hamiltonians $H_s$, $H_e$ and $H_{se}$ for system, environment and their interaction, respectively, and are taken to be
\begin{eqnarray}
   H_{s} &=&J\left(\sum_{j=1}^{l-1} S_j \cdot S_{j+1} + \sum_{j=1}^{l} h_j S^z_j\right) \cr
   H_{e} &=&J\left( \sum_{j=l+1}^{L-1} S_j \cdot S_{j+1} + \sum_{j=l+1}^{L} h_j S^z_j \right)\cr
   H_{se}&=& J( S_l \cdot S_{l+1} + S_1 \cdot S_N)
\end{eqnarray}
where $J$ is the exchange interaction, and the $h_i$ are random fields drawn uniformly in the interval $[-h,+h]$, with $h$ being the disorder strength. The uni-local operator $S_j^\alpha= \sigma_j^\alpha/2$ (for $\alpha=x,y,z$) is the spin operator $\alpha$ at site $j$. The total Hamiltonian is thus given by $H=H_s+H_e+H_{se}$. As the result of this interaction the combined system and environment evolves as $\varrho^{(k)}_{se}(t)=e^{-iHt}\varrho^{(k)}_{se}(0)e^{+iHt}$. By tracing out the environment one can get the reduced density matrix of the system $\varrho_s^{(k)}(t)=\text{Tr}_e\left[ \varrho^{(k)}_{se}(t)\right]$ which also defines our map $\mathcal{E}\left[ \varrho^{(k)} \right] = \varrho_s^{(k)}(t)$. This procedure is shown schematically in Fig.~\ref{fig:fig0}, and its simulation was carried out using the quimb package \cite{gray2018quimb}.
By computing the Holevo quantity in Eq.~(\ref{eq:Holevo_def}) for a given input ensemble $\{p_k,\varrho^{(k)}\}$ and environment state $|e\rangle$ under the action of the map $\mathcal{E}[\cdot]$ one can directly quantify how much information, in bits, can be extracted locally from the system $s$ at time $t$ about its initial state. This is a direct, dynamical quantification of local memory in the subsystem $s$. The value of this quantity in identifying the MBL regime, and probing the ergodic-MBL transition is the subject of the rest of this letter. 

\section{Holevo rate as a quantifier of local memory}
We consider $M{=}2^l$ equiprobable ($p_k = 1/2^l$) messages with $\varrho^{(1)}{=}|0,0,\cdots,0\rangle \langle 0,0,\cdots,0|$ to $\varrho^{(2^l)}{=}|1,1,\cdots,1\rangle \langle 1,1,\cdots,1|$. Three different types of quantum state are taken for the environment: (i) Neel product state $|e_{\text{Neel}}\rangle=|0,1,0,\cdots,1,0\rangle$; (ii) an entangled state resulting from the time evolution of the Neel state under the action of $H_e$, namely $|e_{\text{evo}}\rangle=e^{-iH_et_\text{Neel}}|e_{\text{Neel}}\rangle$ \footnote{We take $t_\text{Neel}=L$ for the entire letter.}; and (iii) one of the mid spectrum eigenstates $|e_{eig}\rangle$ of $H_e$, namely $H_e|e_{eig}\rangle=E_n|e_{eig}\rangle$ where $E_n$ is the median eigenstate energy. For each of these environment types, we compute the Holevo quantity for the given set of equiprobable messages. 
In general, the averaged Holevo quantity $C$ is a function of several variables, namely $C \equiv C(L,l,\{h_j\},t)$ (for a given set of random fields $\{h_j\}$), and is extensive in $l$. As such it is convenient to normalize the Holevo quantity by the message size $l$ to get a \emph{Holevo rate}
\begin{equation}
 R(L,l,\{h_j\},t) = \frac{1}{l}C(L,l,\{h_j\},t). 
\end{equation}
The Holevo rate $R(L,l,\{h_j\},t)$ quantifies what proportion of input data can be extracted at time $t$ by only accessing the qubits in the system $s$; varying between $1$ for perfect memory, and $0$ for full scrambling. We average the Holevo rate over different realizations of the $h_i \in [-h,h]$ for a fixed disorder strength $h$ to get a disorder-averaged Holevo rate $\bar{R}(L,l,h,t)$ \footnote{In this letter, we have used between $100$ and $1000$ samples for each data point to get the disorder-averaged Holevo rate $\bar{R}$.}.
We note here that, additionally to the conventional exponential scaling of computational complexity with total system size $L$, the calculation of the Holevo quantity scales exponentially with the size of the subsystem $l$. This because we need to transmit all $2^l$ messages $\{\rho^{(k)}\}$, which manifests numerically as having to run each disorder sample $2^l$ times. Thus the computational cost of calculating the Holevo quantity for a single disorder realization scales exponentially with subsystem size \emph{in addition} to the standard exponential scaling of the cost of exact diagonalization. For this reason, the system sizes we can access are severely limited; they are doubly afflicted by the `curse of dimensionality'.
\begin{figure*}[ht]
        \includegraphics[width=\linewidth]{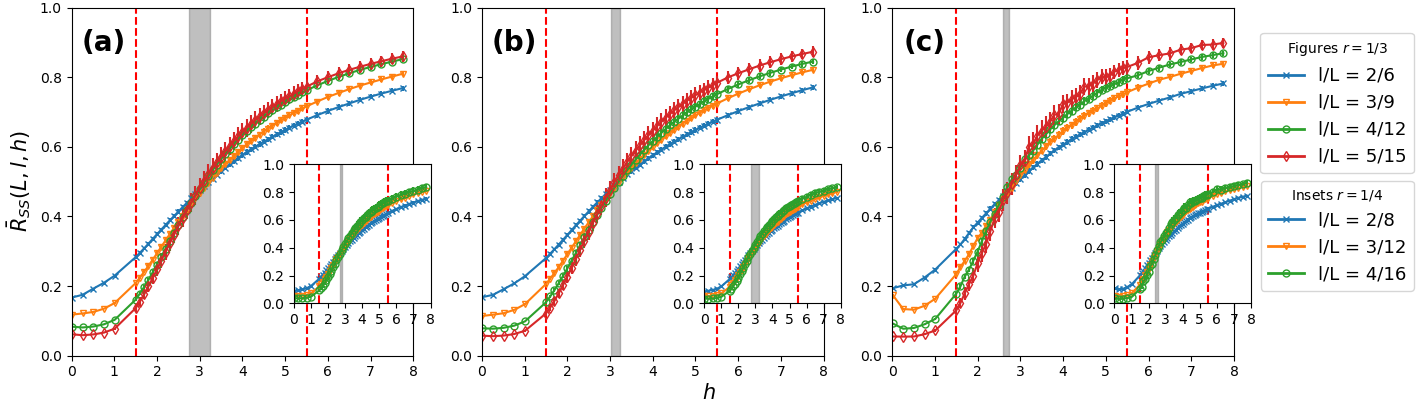}
        \caption{Time-averaged Holevo rate $\bar{R}_{SS}(L,l,h)$ for \textbf{(a)} Neel state, \textbf{(b)} evolved Neel state, and \textbf{(c)} eigenstate environments, respectively. The main figures show results for the fixed ratio $l/L=1/3$, and insets for the fixed ratio $l/L=1/4$. Dashed red lines indicate the region in which the data collapse was carried out, and the grey regions indicate the standard error on $h_c$.}
        \label{fig:fig2}
\end{figure*}
\section{Dynamical Behavior}  To investigate the behavior of the disorder-averaged Holevo rate, in Figs.~\ref{fig:fig1}(a)-(c), we plot  $\bar{R}$  as a function of time for different choices of disorder strength $h$ for $L=16$ and $l=4$ for the three chosen environment states, respectively. After early transient behavior, the disorder-averaged Holevo rate - similarly to other quantities - either saturates rapidly in time (in the ergodic regime) or falls off logarithmically in time and will fully saturate only at exponential time scales (in the MBL regime). In the limit of large $L$ it tends to zero in the ergodic regime, and to a non-zero finite value in the MBL regime. For our finite systems we found that the total evolution times of $T_1 = L^2$ are sufficient to differentiate the two regimes in all cases. The fact that $\bar{R}$ increases as a function of increasing disorder strength indicates that the message subsystem $s$ fails to locally retain information in the ergodic regime, but successfully retains a high proportion of it deep in the MBL regime. In essence, the late-time value of the disorder-averaged Holevo rate successfully captures the conventional understanding of how local memory behaves in both phases.

To estimate the steady-state value of $\bar{R}$, we take the late-time average of the disorder-averaged Holevo rate
\begin{equation}
    \bar{R}_{SS}(L,l,h) = \frac{1}{T_1-T_0}\int^{T_1}_{T_0} \bar{R}(L,l,h,t) \dd t
\end{equation}
In the extreme limit $T_1\to\infty$ this quantity converges to the true steady-state value of $\bar{R}$. Thanks to the short time-scale of transient dynamics in the evolution of $\bar{R}$ the above quantity also closely approximates this value for finite $T_0$, $T_1$ \footnote{We take $T_0 = T_1/8$ for all cases.}, at least to an extent which makes it possible to distinguish ergodic and localized regimes. This time-averaged Holevo rate $\bar{R}_{SS}$ varies between near-zero in the ergodic regime, to near-unity in the fully localized regime; successfully distinguishing regimes. To show this more clearly, in Figs.~\ref{fig:fig1}(d)-(f), we plot $\bar{R}_{SS}$ as a function of disorder strength $h$ for various message sizes $l$ in a chain of size $L=16$ for the three chosen environment states, respectively. As the figures show, $\bar{R}_{SS}$ varies from low to high values as we increase $h$, saturating towards unity. 

Finally, this behavior indicates that the Holevo quantity may exhibit scaling across the MBLT. As the nature of the transition is still under debate, the potential ability of the information-theoretic Holevo quantity to investigate it from the perspective of memory is of great importance. The scaling analysis of the following section addresses this possibility.
\begin{figure*}[ht]
        \includegraphics[width=\linewidth]{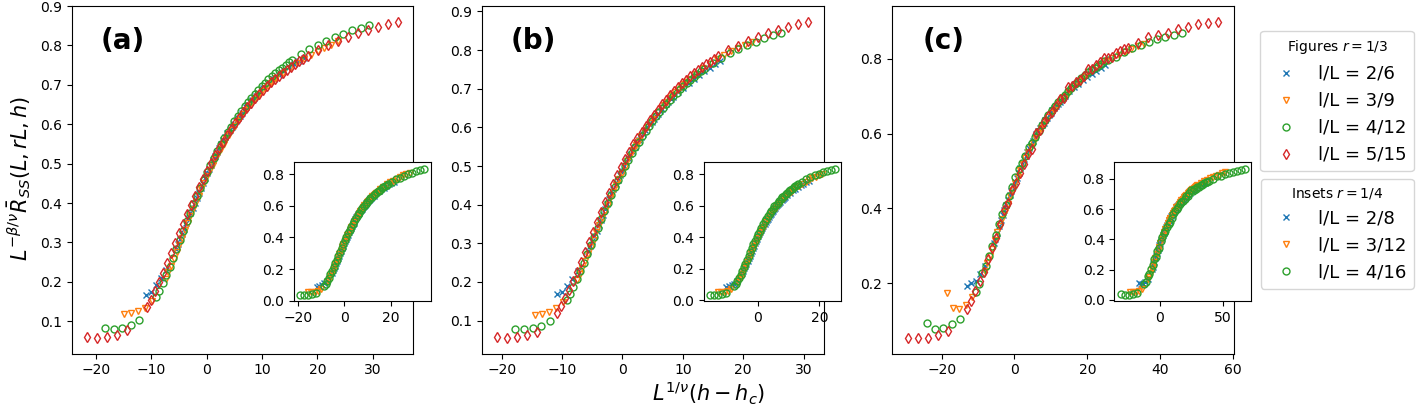}
        \caption{The optimal data collapse of each of the results shown in \rfig{fig2} using the ans\"atze of Eq.~(\ref{eq:finite_size_scaling}) for sizes up to $L=16$. The critical values $h_c$ and exponents $\nu$ and $\beta$ of each collapse are summarized in Table \ref{tab:critical}. }
        \label{fig:fig3}
\end{figure*}
\section{Finite size scaling}
\label{sec:fss}
The behavior of $\bar{R}_{SS}$, presented in Figs.~\ref{fig:fig1}(d)-(f), suggests that $\bar{R}_{SS}$ may show scaling behavior across the MBL transition point. Such a scaling would be invaluable as it would allow us to quantitatively investigate the MBLT from a strictly informational perspective. We note here that such an analysis will exhibit similar pathologies to other extant small-system analyses in the field (see e.g. Ref.~\cite{Devakul2015, Kjall2014}); namely a violation of the Harris criterion \cite{Harris1974, Chayes1986, Chandran2015clo}. This is because, while it is certainly possible to differentiate the two regimes using $\bar{R}_{SS}$, a rigorous analysis of the transition itself is difficult without going to both exponential time scales $T_1\sim\exp{\alpha L}$ and length scales \cite{Panda2020, Sierant2020}. There also exists the ongoing question of the universality class the ergodic-MBL transition falls into \cite{Dumitrescu2019, Morningstar2020, Suntajs2020, Laflorencie2020}. Finally we note an open debate in the field as to whether a stable MBL phase exists at all in the thermodynamic limit or whether it is a finite-size artifact \cite{Suntajs2020, Sels2021, Sierant2020, Crowley2020, Abanin2021}. Despite this, our following scaling analysis, with modest system sizes and times up to $T=L^2$, falls comfortably in line with other extant small-system analyses and validates the Holevo quantity as an information-theoretic counterpart to the quantities more widely used in the field.

In line with previous studies (see e.g. Ref.~\cite{Vosk2015,Gray2018,Pal2010,Khemani2017,Khemani2017a,Luitz2015}), we assume a continuous phase transition with diverging length scale $\xi\sim |h-h_c|^{-\nu}$, where $h_c$ is the infinite-length MBL critical point. The most general two-parameter scaling function for $\bar{R}_{SS}$ can be written as
\begin{equation} \label{eq:finite_size_scaling}
    \bar{R}_{SS}(L,l,h) \sim L^{\beta/\nu} f\left(\frac{l}{L}, L^{1/\nu} (h-h_c)\right)
\end{equation}
where $f(\cdot,\cdot)$ is an arbitrary function and $\beta$ is the exponent that accounts for thermodynamic limit behavior as
$\bar{R}_{SS}(L\rightarrow \infty,l,h) \sim |h-h_c|^\beta f(l/\xi)$. In fact, Eq.~(\ref{eq:finite_size_scaling}) defines a whole family of scaling functions: the functional form of $f(\cdot,\cdot)$ may differ for each environment type and message-to-system length ratio $l/L$.\footnote{This can be seen in the extreme for $l=L$ where $f(\cdot,\cdot)$ is constant in all arguments.}

It is worth emphasizing that, in Eq.~(\ref{eq:finite_size_scaling}), we do not consider a corresponding exponent for the message length $l$ as it is always necessarily constrained by the system length $L$. As such no true thermodynamic limit exists in $l$ independent of the corresponding limit in $L$, and we do not expect to see scaling behavior in $l$ alone. This is evidenced in Figs.~\ref{fig:fig1}(d)-(f), which do not show scaling behavior as we vary $l$ for fixed $L$. For increasing values of $l$, we simply see an overall increase in $\bar{R}_{SS}$ for all $h$; with $\bar{R}_{SS} \to 1$ for all $h$ as $l \to L$.

To verify the scaling ans\"atze in Eq.~(\ref{eq:finite_size_scaling}), in Figs.~\ref{fig:fig2}(a)-(c) we plot $\bar{R}_{SS}$ as a function of $h$ for various choices of $l$ and $L$ such that $l/L$ is fixed. Each panel shows the results for a different type of environment state, namely $|e_{\text{Neel}}\rangle$, $|e_{\text{evo}}\rangle$ and $|e_{\text{eig}}\rangle$, respectively; with the main figures showing fixed $l/L=1/3$, and the insets showing fixed $l/L=1/4$. Interestingly, all the curves in all the three panels and insets intersect at a point, i.e.  $h=h_c$. Demonstrating that, for fixed $l/L$ and a given environment type, $\bar{R}_{SS}$ becomes independent of $L$ and $l$ at $h=h_c$, which means that $\beta \simeq 0$ in all six cases. This indicates that the time-averaged Holevo rate $\bar{R}_{SS}$ is analytic across the transition.

The above scaling analysis provides strong support for the ans\"atze of Eq.~(\ref{eq:finite_size_scaling}) and determines $\beta \simeq 0$. Moreover, the point at which each of these curves intersect can be used to extract values of $h_c$ for a fixed value of $l/L$ and a given environment type. However, it does not provide any estimation for the exponent $\nu$.  
In order to evaluate the critical exponents more directly, we consider another independent finite size scaling analysis using the Python package pyfssa \cite{sorge2015pyfssa,melchert2009autoscalepy}. In Figs.~\ref{fig:fig3}(a)-(c) we plot $L^{-\beta/\nu}\bar{R}_{SS}$ as a function of $L^{1/\nu}(h-h_c)$ for various choices of $l$ and $L$ while keeping $l/L$ fixed for the three given environment states, respectively. By properly tuning the critical value $h_c$, and the exponents $\beta$ and $\nu$ one can get a separate data collapse for each set of curves. As the figures show, different critical parameters are obtained for each environment. Interestingly, in all cases the exponent $\beta$ is very small which is consistent with the previous scaling analysis. The results of these data collapses in the form of extracted critical values and exponents are summarized in Table \ref{tab:critical}. We find that the extracted values are consistent between different message-to-system length ratios but vary as the environment type changes.


\begin{table}[ht]
\addvbuffer[10pt 0pt]{
\begin{tabular}{|c|c|ccc|}
\hline
$l/L$ & Environment & $h_c$ & $\nu$ & $\beta$ \\
\hline
\multirow{3}{*}{1/3} & Neel & $3.26\pm 0.18$ & $1.32\pm0.27$ & $0.00$ \\
 & Evolved Neel & $3.41\pm 0.11$ & $1.40\pm0.14$ & $0.00$ \\
 & Eigenstate & $2.87 \pm 0.05$ & $1.06\pm0.09$ & $0.00$ \\
 \hline
 \multirow{3}{*}{1/4} & Neel & $3.07\pm 0.06$ & $1.38\pm0.10$ & $0.00$ \\
 & Evolved Neel & $3.26\pm 0.20$ & $1.57\pm0.26$ & $0.00$ \\
 & Eigenstate & $2.69 \pm 0.10$ & $1.04\pm0.13$ & $0.00$ \\
 \hline
 \end{tabular} }
 \caption{\textbf{\label{tab:table-name}} Table of extracted critical values and exponents for all the investigated message-to-system length ratios and environment types. The standard error on $\beta$ was on the order of $0.01$ or less for all results, and is omitted in the above table. }
 \label{tab:critical}
 \vspace{-1em}
\end{table}
\section{Role of the environment} A noteworthy feature of Table~\ref{tab:critical} is that the values of $h_c$ and $\nu$ vary between environments. This is expected for the critical value $h_c$, as each environment makes the state of the system overlap with different regions of the MBL mobility edge. Interestingly, our work indicates a similar variation in $\nu$. We conjecture two possible explanations for this variation: (i) that the value of the critical exponent $\nu$ might also vary across the MBL mobility edge; or (ii) that $\nu$ in the thermodynamic limit is unique, but different environments may replicate the behavior of the system at the thermodynamic limit better than the others \footnote{This latter suggestion is motivated by evidence that small-scale numerics violate the Harris criterion which insist that $\nu > 2$ \cite{Harris1974, Chayes1986, Chandran2015clo}. For example, in Table~\ref{tab:critical}, the eigenstate environment types violate this bound the most, which may indicate that they are more susceptible to finite-size effects than the other two; driving down the value of $\nu$.}.





\section{Conclusions} We have introduced the dynamical Holevo quantity as a complete and concrete quantifier of local memory, in terms of numbers of preserved bits. After a discussion of a wide range of extant quantities and their varied shortcomings when it comes to actually quantifying memory, we have argued that a strictly informational approach - and the Holevo quantity in particular - is the most complete way to access local memory in these systems. We have shown that the Holevo quantity can successfully distinguish ergodic and localized regimes and also exhibits scaling behavior across the MBLT. In particular, we have determined a family of two-parameter scaling ans\"atze for the steady-state from which we extract critical values and exponents in line with these extant numerics for modest system sizes and time scales. The results of this letter place the concept of local memory across the MBLT on a clear quantitative footing; and is, to our knowledge, the first quantitative investigation of local memory from a strictly informational perspective in \emph{any} quantum many-body system.






\section{Acknowledgements} AB acknowledges the National Key R\&D Program of China, Grant No.2018YFA0306703. SB and AB acknowledge the
EPSRC grant Nonergodic quantum manipulation
EP/R029075/1. ANK acknowledges support from the EPSRC.


%


\begin{thebibliography}{93}%
\makeatletter
\providecommand \@ifxundefined [1]{%
 \@ifx{#1\undefined}
}%
\providecommand \@ifnum [1]{%
 \ifnum #1\expandafter \@firstoftwo
 \else \expandafter \@secondoftwo
 \fi
}%
\providecommand \@ifx [1]{%
 \ifx #1\expandafter \@firstoftwo
 \else \expandafter \@secondoftwo
 \fi
}%
\providecommand \natexlab [1]{#1}%
\providecommand \enquote  [1]{``#1''}%
\providecommand \bibnamefont  [1]{#1}%
\providecommand \bibfnamefont [1]{#1}%
\providecommand \citenamefont [1]{#1}%
\providecommand \href@noop [0]{\@secondoftwo}%
\providecommand \href [0]{\begingroup \@sanitize@url \@href}%
\providecommand \@href[1]{\@@startlink{#1}\@@href}%
\providecommand \@@href[1]{\endgroup#1\@@endlink}%
\providecommand \@sanitize@url [0]{\catcode `\\12\catcode `\$12\catcode
  `\&12\catcode `\#12\catcode `\^12\catcode `\_12\catcode `\%12\relax}%
\providecommand \@@startlink[1]{}%
\providecommand \@@endlink[0]{}%
\providecommand \url  [0]{\begingroup\@sanitize@url \@url }%
\providecommand \@url [1]{\endgroup\@href {#1}{\urlprefix }}%
\providecommand \urlprefix  [0]{URL }%
\providecommand \Eprint [0]{\href }%
\providecommand \doibase [0]{http://dx.doi.org/}%
\providecommand \selectlanguage [0]{\@gobble}%
\providecommand \bibinfo  [0]{\@secondoftwo}%
\providecommand \bibfield  [0]{\@secondoftwo}%
\providecommand \translation [1]{[#1]}%
\providecommand \BibitemOpen [0]{}%
\providecommand \bibitemStop [0]{}%
\providecommand \bibitemNoStop [0]{.\EOS\space}%
\providecommand \EOS [0]{\spacefactor3000\relax}%
\providecommand \BibitemShut  [1]{\csname bibitem#1\endcsname}%
\let\auto@bib@innerbib\@empty
\bibitem [{\citenamefont {Nandkishore}\ and\ \citenamefont
  {Huse}(2015)}]{Nandkishore2015}%
  \BibitemOpen
  \bibfield  {author} {\bibinfo {author} {\bibfnamefont {R.}~\bibnamefont
  {Nandkishore}}\ and\ \bibinfo {author} {\bibfnamefont {D.~A.}\ \bibnamefont
  {Huse}},\ }\href {\doibase 10.1146/annurev-conmatphys-031214-014726}
  {\bibfield  {journal} {\bibinfo  {journal} {Annual Review of Condensed Matter
  Physics}\ }\textbf {\bibinfo {volume} {6}},\ \bibinfo {pages} {15} (\bibinfo
  {year} {2015})}\BibitemShut {NoStop}%
\bibitem [{\citenamefont {Abanin}\ \emph {et~al.}(2019)\citenamefont {Abanin},
  \citenamefont {Altman}, \citenamefont {Bloch},\ and\ \citenamefont
  {Serbyn}}]{Abanin2019}%
  \BibitemOpen
  \bibfield  {author} {\bibinfo {author} {\bibfnamefont {D.~A.}\ \bibnamefont
  {Abanin}}, \bibinfo {author} {\bibfnamefont {E.}~\bibnamefont {Altman}},
  \bibinfo {author} {\bibfnamefont {I.}~\bibnamefont {Bloch}}, \ and\ \bibinfo
  {author} {\bibfnamefont {M.}~\bibnamefont {Serbyn}},\ }\href {\doibase
  10.1103/RevModPhys.91.021001} {\bibfield  {journal} {\bibinfo  {journal}
  {Rev. Mod. Phys.}\ }\textbf {\bibinfo {volume} {91}},\ \bibinfo {pages}
  {021001} (\bibinfo {year} {2019})}\BibitemShut {NoStop}%
\bibitem [{\citenamefont {Alet}\ and\ \citenamefont
  {Laflorencie}(2018)}]{Alet2018}%
  \BibitemOpen
  \bibfield  {author} {\bibinfo {author} {\bibfnamefont {F.}~\bibnamefont
  {Alet}}\ and\ \bibinfo {author} {\bibfnamefont {N.}~\bibnamefont
  {Laflorencie}},\ }\href {\doibase 10.1016/j.crhy.2018.03.003} {\bibfield
  {journal} {\bibinfo  {journal} {Comptes Rendus Physique}\ }\textbf {\bibinfo
  {volume} {19}},\ \bibinfo {pages} {498} (\bibinfo {year} {2018})}\BibitemShut
  {NoStop}%
\bibitem [{\citenamefont {Sachdev}(2011)}]{sachdev2011}%
  \BibitemOpen
  \bibfield  {author} {\bibinfo {author} {\bibfnamefont {S.}~\bibnamefont
  {Sachdev}},\ }\href@noop {} {\emph {\bibinfo {title} {Quantum phase
  transitions}}},\ \bibinfo {edition} {2nd}\ ed.\ (\bibinfo  {publisher}
  {Cambridge University Press},\ \bibinfo {address} {Cambridge},\ \bibinfo
  {year} {2011})\BibitemShut {NoStop}%
\bibitem [{\citenamefont {Abanin}\ and\ \citenamefont {Papi{\'{c}
  }}(2017)}]{Abanin2017}%
  \BibitemOpen
  \bibfield  {author} {\bibinfo {author} {\bibfnamefont {D.~A.}\ \bibnamefont
  {Abanin}}\ and\ \bibinfo {author} {\bibfnamefont {Z.}~\bibnamefont
  {Papi{\'{c} }}},\ }\href {\doibase 10.1002/andp.201700169} {\bibfield
  {journal} {\bibinfo  {journal} {Annalen der Physik}\ }\textbf {\bibinfo
  {volume} {529}},\ \bibinfo {pages} {1700169} (\bibinfo {year}
  {2017})}\BibitemShut {NoStop}%
\bibitem [{\citenamefont {Oganesyan}\ and\ \citenamefont
  {Huse}(2007)}]{Oganesyan2007}%
  \BibitemOpen
  \bibfield  {author} {\bibinfo {author} {\bibfnamefont {V.}~\bibnamefont
  {Oganesyan}}\ and\ \bibinfo {author} {\bibfnamefont {D.~A.}\ \bibnamefont
  {Huse}},\ }\href {\doibase 10.1103/PhysRevB.75.155111} {\bibfield  {journal}
  {\bibinfo  {journal} {Phys. Rev. B}\ }\textbf {\bibinfo {volume} {75}},\
  \bibinfo {pages} {155111} (\bibinfo {year} {2007})}\BibitemShut {NoStop}%
\bibitem [{\citenamefont {Pal}\ and\ \citenamefont {Huse}(2010)}]{Pal2010}%
  \BibitemOpen
  \bibfield  {author} {\bibinfo {author} {\bibfnamefont {A.}~\bibnamefont
  {Pal}}\ and\ \bibinfo {author} {\bibfnamefont {D.~A.}\ \bibnamefont {Huse}},\
  }\href@noop {} {\bibfield  {journal} {\bibinfo  {journal} {Phys. Rev. B}\
  }\textbf {\bibinfo {volume} {82}},\ \bibinfo {pages} {174411} (\bibinfo
  {year} {2010})}\BibitemShut {NoStop}%
\bibitem [{\citenamefont {Kj{\"{a}}ll}(2018)}]{Kjall2018}%
  \BibitemOpen
  \bibfield  {author} {\bibinfo {author} {\bibfnamefont {J.~A.}\ \bibnamefont
  {Kj{\"{a}}ll}},\ }\href@noop {} {\bibfield  {journal} {\bibinfo  {journal}
  {Phys. Rev. B}\ }\textbf {\bibinfo {volume} {97}},\ \bibinfo {pages} {035163}
  (\bibinfo {year} {2018})}\BibitemShut {NoStop}%
\bibitem [{\citenamefont {Sierant}\ and\ \citenamefont
  {Zakrzewski}(2017)}]{Sierant2017}%
  \BibitemOpen
  \bibfield  {author} {\bibinfo {author} {\bibfnamefont {P.}~\bibnamefont
  {Sierant}}\ and\ \bibinfo {author} {\bibfnamefont {J.}~\bibnamefont
  {Zakrzewski}},\ }\href@noop {} {\bibfield  {journal} {\bibinfo  {journal}
  {New J. Phys.}\ }\textbf {\bibinfo {volume} {20}},\ \bibinfo {pages} {043032}
  (\bibinfo {year} {2017})}\BibitemShut {NoStop}%
\bibitem [{\citenamefont {Roushan}\ \emph {et~al.}(2017)\citenamefont
  {Roushan}, \citenamefont {Neill}, \citenamefont {Tangpanitanon},
  \citenamefont {Bastidas}, \citenamefont {Megrant}, \citenamefont {Barends},
  \citenamefont {Chen}, \citenamefont {Chen}, \citenamefont {Chiaro},
  \citenamefont {Dunsworth}, \citenamefont {Fowler}, \citenamefont {Foxen},
  \citenamefont {Giustina}, \citenamefont {Jeffrey}, \citenamefont {Kelly},
  \citenamefont {Lucero}, \citenamefont {Mutus}, \citenamefont {Neeley},
  \citenamefont {Quintana}, \citenamefont {Sank}, \citenamefont {Vainsencher},
  \citenamefont {Wenner}, \citenamefont {White}, \citenamefont {Neven},
  \citenamefont {Angelakis},\ and\ \citenamefont {Martinis}}]{Roushan2017}%
  \BibitemOpen
  \bibfield  {author} {\bibinfo {author} {\bibfnamefont {P.}~\bibnamefont
  {Roushan}}, \bibinfo {author} {\bibfnamefont {C.}~\bibnamefont {Neill}},
  \bibinfo {author} {\bibfnamefont {J.}~\bibnamefont {Tangpanitanon}}, \bibinfo
  {author} {\bibfnamefont {V.~M.}\ \bibnamefont {Bastidas}}, \bibinfo {author}
  {\bibfnamefont {A.}~\bibnamefont {Megrant}}, \bibinfo {author} {\bibfnamefont
  {R.}~\bibnamefont {Barends}}, \bibinfo {author} {\bibfnamefont
  {Y.}~\bibnamefont {Chen}}, \bibinfo {author} {\bibfnamefont {Z.}~\bibnamefont
  {Chen}}, \bibinfo {author} {\bibfnamefont {B.}~\bibnamefont {Chiaro}},
  \bibinfo {author} {\bibfnamefont {A.}~\bibnamefont {Dunsworth}}, \bibinfo
  {author} {\bibfnamefont {A.}~\bibnamefont {Fowler}}, \bibinfo {author}
  {\bibfnamefont {B.}~\bibnamefont {Foxen}}, \bibinfo {author} {\bibfnamefont
  {M.}~\bibnamefont {Giustina}}, \bibinfo {author} {\bibfnamefont
  {E.}~\bibnamefont {Jeffrey}}, \bibinfo {author} {\bibfnamefont
  {J.}~\bibnamefont {Kelly}}, \bibinfo {author} {\bibfnamefont
  {E.}~\bibnamefont {Lucero}}, \bibinfo {author} {\bibfnamefont
  {J.}~\bibnamefont {Mutus}}, \bibinfo {author} {\bibfnamefont
  {M.}~\bibnamefont {Neeley}}, \bibinfo {author} {\bibfnamefont
  {C.}~\bibnamefont {Quintana}}, \bibinfo {author} {\bibfnamefont
  {D.}~\bibnamefont {Sank}}, \bibinfo {author} {\bibfnamefont {A.}~\bibnamefont
  {Vainsencher}}, \bibinfo {author} {\bibfnamefont {J.}~\bibnamefont {Wenner}},
  \bibinfo {author} {\bibfnamefont {T.}~\bibnamefont {White}}, \bibinfo
  {author} {\bibfnamefont {H.}~\bibnamefont {Neven}}, \bibinfo {author}
  {\bibfnamefont {D.~G.}\ \bibnamefont {Angelakis}}, \ and\ \bibinfo {author}
  {\bibfnamefont {J.}~\bibnamefont {Martinis}},\ }\href {\doibase
  10.1126/science.aao1401} {\bibfield  {journal} {\bibinfo  {journal}
  {Science}\ }\textbf {\bibinfo {volume} {358}},\ \bibinfo {pages} {1175}
  (\bibinfo {year} {2017})}\BibitemShut {NoStop}%
\bibitem [{\citenamefont {Song}\ and\ \citenamefont
  {Shepelyansky}(2000)}]{Song1999}%
  \BibitemOpen
  \bibfield  {author} {\bibinfo {author} {\bibfnamefont {P.~H.}\ \bibnamefont
  {Song}}\ and\ \bibinfo {author} {\bibfnamefont {D.~L.}\ \bibnamefont
  {Shepelyansky}},\ }\href {\doibase 10.1103/PhysRevB.61.15546} {\bibfield
  {journal} {\bibinfo  {journal} {Phys. Rev. B}\ }\textbf {\bibinfo {volume}
  {61}},\ \bibinfo {pages} {15546} (\bibinfo {year} {2000})}\BibitemShut
  {NoStop}%
\bibitem [{\citenamefont {Eisert}\ \emph {et~al.}(2010)\citenamefont {Eisert},
  \citenamefont {Cramer},\ and\ \citenamefont {Plenio}}]{Eisert2010}%
  \BibitemOpen
  \bibfield  {author} {\bibinfo {author} {\bibfnamefont {J.}~\bibnamefont
  {Eisert}}, \bibinfo {author} {\bibfnamefont {M.}~\bibnamefont {Cramer}}, \
  and\ \bibinfo {author} {\bibfnamefont {M.~B.}\ \bibnamefont {Plenio}},\
  }\href {\doibase 10.1103/revmodphys.82.277} {\bibfield  {journal} {\bibinfo
  {journal} {Reviews of Modern Physics}\ }\textbf {\bibinfo {volume} {82}},\
  \bibinfo {pages} {277} (\bibinfo {year} {2010})}\BibitemShut {NoStop}%
\bibitem [{\citenamefont {Grover}(2014)}]{Grover2014}%
  \BibitemOpen
  \bibfield  {author} {\bibinfo {author} {\bibfnamefont {T.}~\bibnamefont
  {Grover}},\ }\href {\doibase 10.48550/ARXIV.1405.1471} {\enquote {\bibinfo
  {title} {Certain general constraints on the many-body localization
  transition},}\ } (\bibinfo {year} {2014})\BibitemShut {NoStop}%
\bibitem [{\citenamefont {Bauer}\ and\ \citenamefont
  {Nayak}(2013)}]{Bauer2013}%
  \BibitemOpen
  \bibfield  {author} {\bibinfo {author} {\bibfnamefont {B.}~\bibnamefont
  {Bauer}}\ and\ \bibinfo {author} {\bibfnamefont {C.}~\bibnamefont {Nayak}},\
  }\href@noop {} {\bibfield  {journal} {\bibinfo  {journal} {J. Stat. Mech.:
  Theory Exp.}\ }\textbf {\bibinfo {volume} {2013}},\ \bibinfo {pages} {09005}
  (\bibinfo {year} {2013})}\BibitemShut {NoStop}%
\bibitem [{\citenamefont {Kj\"all}\ \emph {et~al.}(2014)\citenamefont
  {Kj\"all}, \citenamefont {Bardarson},\ and\ \citenamefont
  {Pollmann}}]{Kjall2014}%
  \BibitemOpen
  \bibfield  {author} {\bibinfo {author} {\bibfnamefont {J.~A.}\ \bibnamefont
  {Kj\"all}}, \bibinfo {author} {\bibfnamefont {J.~H.}\ \bibnamefont
  {Bardarson}}, \ and\ \bibinfo {author} {\bibfnamefont {F.}~\bibnamefont
  {Pollmann}},\ }\href {\doibase 10.1103/PhysRevLett.113.107204} {\bibfield
  {journal} {\bibinfo  {journal} {Phys. Rev. Lett.}\ }\textbf {\bibinfo
  {volume} {113}},\ \bibinfo {pages} {107204} (\bibinfo {year}
  {2014})}\BibitemShut {NoStop}%
\bibitem [{\citenamefont {Khemani}\ \emph
  {et~al.}(2017{\natexlab{a}})\citenamefont {Khemani}, \citenamefont {Lim},
  \citenamefont {Sheng},\ and\ \citenamefont {Huse}}]{Khemani2017}%
  \BibitemOpen
  \bibfield  {author} {\bibinfo {author} {\bibfnamefont {V.}~\bibnamefont
  {Khemani}}, \bibinfo {author} {\bibfnamefont {S.~P.}\ \bibnamefont {Lim}},
  \bibinfo {author} {\bibfnamefont {D.~N.}\ \bibnamefont {Sheng}}, \ and\
  \bibinfo {author} {\bibfnamefont {D.~A.}\ \bibnamefont {Huse}},\ }\href@noop
  {} {\bibfield  {journal} {\bibinfo  {journal} {Phys. Rev. X}\ }\textbf
  {\bibinfo {volume} {7}},\ \bibinfo {pages} {021013} (\bibinfo {year}
  {2017}{\natexlab{a}})}\BibitemShut {NoStop}%
\bibitem [{\citenamefont {Khemani}\ \emph
  {et~al.}(2017{\natexlab{b}})\citenamefont {Khemani}, \citenamefont {Sheng},\
  and\ \citenamefont {Huse}}]{Khemani2017a}%
  \BibitemOpen
  \bibfield  {author} {\bibinfo {author} {\bibfnamefont {V.}~\bibnamefont
  {Khemani}}, \bibinfo {author} {\bibfnamefont {D.~N.}\ \bibnamefont {Sheng}},
  \ and\ \bibinfo {author} {\bibfnamefont {D.~A.}\ \bibnamefont {Huse}},\
  }\href {\doibase 10.1103/PhysRevLett.119.075702} {\bibfield  {journal}
  {\bibinfo  {journal} {Phys. Rev. Lett.}\ }\textbf {\bibinfo {volume} {119}},\
  \bibinfo {pages} {075702} (\bibinfo {year} {2017}{\natexlab{b}})}\BibitemShut
  {NoStop}%
\bibitem [{\citenamefont {Lukin}\ \emph {et~al.}(2019)\citenamefont {Lukin},
  \citenamefont {Rispoli}, \citenamefont {Schittko}, \citenamefont {Tai},
  \citenamefont {Kaufman}, \citenamefont {Choi}, \citenamefont {Khemani},
  \citenamefont {L{\'{e} }onard},\ and\ \citenamefont {Greiner}}]{Lukin2019}%
  \BibitemOpen
  \bibfield  {author} {\bibinfo {author} {\bibfnamefont {A.}~\bibnamefont
  {Lukin}}, \bibinfo {author} {\bibfnamefont {M.}~\bibnamefont {Rispoli}},
  \bibinfo {author} {\bibfnamefont {R.}~\bibnamefont {Schittko}}, \bibinfo
  {author} {\bibfnamefont {M.~E.}\ \bibnamefont {Tai}}, \bibinfo {author}
  {\bibfnamefont {A.~M.}\ \bibnamefont {Kaufman}}, \bibinfo {author}
  {\bibfnamefont {S.}~\bibnamefont {Choi}}, \bibinfo {author} {\bibfnamefont
  {V.}~\bibnamefont {Khemani}}, \bibinfo {author} {\bibfnamefont
  {J.}~\bibnamefont {L{\'{e} }onard}}, \ and\ \bibinfo {author} {\bibfnamefont
  {M.}~\bibnamefont {Greiner}},\ }\href {\doibase 10.1126/science.aau0818}
  {\bibfield  {journal} {\bibinfo  {journal} {Science}\ }\textbf {\bibinfo
  {volume} {364}},\ \bibinfo {pages} {256} (\bibinfo {year}
  {2019})}\BibitemShut {NoStop}%
\bibitem [{\citenamefont {Wei}\ \emph {et~al.}(2018)\citenamefont {Wei},
  \citenamefont {Ramanathan},\ and\ \citenamefont {Cappellaro}}]{Wei2018}%
  \BibitemOpen
  \bibfield  {author} {\bibinfo {author} {\bibfnamefont {K.~X.}\ \bibnamefont
  {Wei}}, \bibinfo {author} {\bibfnamefont {C.}~\bibnamefont {Ramanathan}}, \
  and\ \bibinfo {author} {\bibfnamefont {P.}~\bibnamefont {Cappellaro}},\
  }\href@noop {} {\bibfield  {journal} {\bibinfo  {journal} {Phys. Rev. Lett.}\
  }\textbf {\bibinfo {volume} {120}},\ \bibinfo {pages} {070501} (\bibinfo
  {year} {2018})}\BibitemShut {NoStop}%
\bibitem [{\citenamefont {{Bar Lev}}\ \emph {et~al.}(2015)\citenamefont {{Bar
  Lev}}, \citenamefont {Cohen},\ and\ \citenamefont {Reichman}}]{BarLev2015}%
  \BibitemOpen
  \bibfield  {author} {\bibinfo {author} {\bibfnamefont {Y.}~\bibnamefont {{Bar
  Lev}}}, \bibinfo {author} {\bibfnamefont {G.}~\bibnamefont {Cohen}}, \ and\
  \bibinfo {author} {\bibfnamefont {D.~R.}\ \bibnamefont {Reichman}},\
  }\href@noop {} {\bibfield  {journal} {\bibinfo  {journal} {Phys. Rev. Lett.}\
  }\textbf {\bibinfo {volume} {114}},\ \bibinfo {pages} {100601} (\bibinfo
  {year} {2015})}\BibitemShut {NoStop}%
\bibitem [{\citenamefont {Iemini}\ \emph {et~al.}(2016)\citenamefont {Iemini},
  \citenamefont {Russomanno}, \citenamefont {Rossini}, \citenamefont
  {Scardicchio},\ and\ \citenamefont {Fazio}}]{Iemini2016}%
  \BibitemOpen
  \bibfield  {author} {\bibinfo {author} {\bibfnamefont {F.}~\bibnamefont
  {Iemini}}, \bibinfo {author} {\bibfnamefont {A.}~\bibnamefont {Russomanno}},
  \bibinfo {author} {\bibfnamefont {D.}~\bibnamefont {Rossini}}, \bibinfo
  {author} {\bibfnamefont {A.}~\bibnamefont {Scardicchio}}, \ and\ \bibinfo
  {author} {\bibfnamefont {R.}~\bibnamefont {Fazio}},\ }\href@noop {}
  {\bibfield  {journal} {\bibinfo  {journal} {Phys. Rev. B}\ }\textbf {\bibinfo
  {volume} {94}},\ \bibinfo {pages} {214206} (\bibinfo {year}
  {2016})}\BibitemShut {NoStop}%
\bibitem [{\citenamefont {Lezama}\ \emph {et~al.}(2017)\citenamefont {Lezama},
  \citenamefont {Bera}, \citenamefont {Schomerus}, \citenamefont
  {Heidrich-Meisner},\ and\ \citenamefont {Bardarson}}]{Lezama2017}%
  \BibitemOpen
  \bibfield  {author} {\bibinfo {author} {\bibfnamefont {T.~L.}\ \bibnamefont
  {Lezama}}, \bibinfo {author} {\bibfnamefont {S.}~\bibnamefont {Bera}},
  \bibinfo {author} {\bibfnamefont {H.}~\bibnamefont {Schomerus}}, \bibinfo
  {author} {\bibfnamefont {F.}~\bibnamefont {Heidrich-Meisner}}, \ and\
  \bibinfo {author} {\bibfnamefont {J.~H.}\ \bibnamefont {Bardarson}},\
  }\href@noop {} {\bibfield  {journal} {\bibinfo  {journal} {Phys. Rev. B}\
  }\textbf {\bibinfo {volume} {96}},\ \bibinfo {pages} {060202} (\bibinfo
  {year} {2017})}\BibitemShut {NoStop}%
\bibitem [{\citenamefont {Schreiber}\ \emph {et~al.}(2015)\citenamefont
  {Schreiber}, \citenamefont {Hodgman}, \citenamefont {Bordia}, \citenamefont
  {Lüschen}, \citenamefont {Fischer}, \citenamefont {Vosk}, \citenamefont
  {Altman}, \citenamefont {Schneider},\ and\ \citenamefont
  {Bloch}}]{Schreiber2015}%
  \BibitemOpen
  \bibfield  {author} {\bibinfo {author} {\bibfnamefont {M.}~\bibnamefont
  {Schreiber}}, \bibinfo {author} {\bibfnamefont {S.~S.}\ \bibnamefont
  {Hodgman}}, \bibinfo {author} {\bibfnamefont {P.}~\bibnamefont {Bordia}},
  \bibinfo {author} {\bibfnamefont {H.~P.}\ \bibnamefont {Lüschen}}, \bibinfo
  {author} {\bibfnamefont {M.~H.}\ \bibnamefont {Fischer}}, \bibinfo {author}
  {\bibfnamefont {R.}~\bibnamefont {Vosk}}, \bibinfo {author} {\bibfnamefont
  {E.}~\bibnamefont {Altman}}, \bibinfo {author} {\bibfnamefont
  {U.}~\bibnamefont {Schneider}}, \ and\ \bibinfo {author} {\bibfnamefont
  {I.}~\bibnamefont {Bloch}},\ }\href {\doibase 10.1126/science.aaa7432}
  {\bibfield  {journal} {\bibinfo  {journal} {Science}\ }\textbf {\bibinfo
  {volume} {349}},\ \bibinfo {pages} {842} (\bibinfo {year}
  {2015})}\BibitemShut {NoStop}%
\bibitem [{\citenamefont {yoon Choi}\ \emph {et~al.}(2016)\citenamefont {yoon
  Choi}, \citenamefont {Hild}, \citenamefont {Zeiher}, \citenamefont
  {Schau{\ss}}, \citenamefont {Rubio-Abadal}, \citenamefont {Yefsah},
  \citenamefont {Khemani}, \citenamefont {Huse}, \citenamefont {Bloch},\ and\
  \citenamefont {Gross}}]{Choi2016}%
  \BibitemOpen
  \bibfield  {author} {\bibinfo {author} {\bibfnamefont {J.}~\bibnamefont {yoon
  Choi}}, \bibinfo {author} {\bibfnamefont {S.}~\bibnamefont {Hild}}, \bibinfo
  {author} {\bibfnamefont {J.}~\bibnamefont {Zeiher}}, \bibinfo {author}
  {\bibfnamefont {P.}~\bibnamefont {Schau{\ss}}}, \bibinfo {author}
  {\bibfnamefont {A.}~\bibnamefont {Rubio-Abadal}}, \bibinfo {author}
  {\bibfnamefont {T.}~\bibnamefont {Yefsah}}, \bibinfo {author} {\bibfnamefont
  {V.}~\bibnamefont {Khemani}}, \bibinfo {author} {\bibfnamefont {D.~A.}\
  \bibnamefont {Huse}}, \bibinfo {author} {\bibfnamefont {I.}~\bibnamefont
  {Bloch}}, \ and\ \bibinfo {author} {\bibfnamefont {C.}~\bibnamefont
  {Gross}},\ }\href {\doibase 10.1126/science.aaf8834} {\bibfield  {journal}
  {\bibinfo  {journal} {Science}\ }\textbf {\bibinfo {volume} {352}},\ \bibinfo
  {pages} {1547} (\bibinfo {year} {2016})}\BibitemShut {NoStop}%
\bibitem [{\citenamefont {Rubio-Abadal}\ \emph {et~al.}(2019)\citenamefont
  {Rubio-Abadal}, \citenamefont {Choi}, \citenamefont {Zeiher}, \citenamefont
  {Hollerith}, \citenamefont {Rui}, \citenamefont {Bloch},\ and\ \citenamefont
  {Gross}}]{Rubio-Abadal2019}%
  \BibitemOpen
  \bibfield  {author} {\bibinfo {author} {\bibfnamefont {A.}~\bibnamefont
  {Rubio-Abadal}}, \bibinfo {author} {\bibfnamefont {J.~Y.}\ \bibnamefont
  {Choi}}, \bibinfo {author} {\bibfnamefont {J.}~\bibnamefont {Zeiher}},
  \bibinfo {author} {\bibfnamefont {S.}~\bibnamefont {Hollerith}}, \bibinfo
  {author} {\bibfnamefont {J.}~\bibnamefont {Rui}}, \bibinfo {author}
  {\bibfnamefont {I.}~\bibnamefont {Bloch}}, \ and\ \bibinfo {author}
  {\bibfnamefont {C.}~\bibnamefont {Gross}},\ }\href@noop {} {\bibfield
  {journal} {\bibinfo  {journal} {Phys. Rev. X}\ }\textbf {\bibinfo {volume}
  {9}},\ \bibinfo {pages} {041014} (\bibinfo {year} {2019})}\BibitemShut
  {NoStop}%
\bibitem [{\citenamefont {Xu}\ \emph {et~al.}(2018)\citenamefont {Xu},
  \citenamefont {Chen}, \citenamefont {Zeng}, \citenamefont {Zhang},
  \citenamefont {Song}, \citenamefont {Liu}, \citenamefont {Guo}, \citenamefont
  {Zhang}, \citenamefont {Xu}, \citenamefont {Deng}, \citenamefont {Huang},
  \citenamefont {Wang}, \citenamefont {Zhu}, \citenamefont {Zheng},\ and\
  \citenamefont {Fan}}]{Xu2018}%
  \BibitemOpen
  \bibfield  {author} {\bibinfo {author} {\bibfnamefont {K.}~\bibnamefont
  {Xu}}, \bibinfo {author} {\bibfnamefont {J.~J.}\ \bibnamefont {Chen}},
  \bibinfo {author} {\bibfnamefont {Y.}~\bibnamefont {Zeng}}, \bibinfo {author}
  {\bibfnamefont {Y.~R.}\ \bibnamefont {Zhang}}, \bibinfo {author}
  {\bibfnamefont {C.}~\bibnamefont {Song}}, \bibinfo {author} {\bibfnamefont
  {W.}~\bibnamefont {Liu}}, \bibinfo {author} {\bibfnamefont {Q.}~\bibnamefont
  {Guo}}, \bibinfo {author} {\bibfnamefont {P.}~\bibnamefont {Zhang}}, \bibinfo
  {author} {\bibfnamefont {D.}~\bibnamefont {Xu}}, \bibinfo {author}
  {\bibfnamefont {H.}~\bibnamefont {Deng}}, \bibinfo {author} {\bibfnamefont
  {K.}~\bibnamefont {Huang}}, \bibinfo {author} {\bibfnamefont
  {H.}~\bibnamefont {Wang}}, \bibinfo {author} {\bibfnamefont {X.}~\bibnamefont
  {Zhu}}, \bibinfo {author} {\bibfnamefont {D.}~\bibnamefont {Zheng}}, \ and\
  \bibinfo {author} {\bibfnamefont {H.}~\bibnamefont {Fan}},\ }\href@noop {}
  {\bibfield  {journal} {\bibinfo  {journal} {Phys. Rev. Lett.}\ }\textbf
  {\bibinfo {volume} {120}},\ \bibinfo {pages} {050507} (\bibinfo {year}
  {2018})}\BibitemShut {NoStop}%
\bibitem [{\citenamefont {Bordia}\ \emph {et~al.}(2017)\citenamefont {Bordia},
  \citenamefont {L{\"{u}}schen}, \citenamefont {Scherg}, \citenamefont
  {Gopalakrishnan}, \citenamefont {Knap}, \citenamefont {Schneider},\ and\
  \citenamefont {Bloch}}]{Bordia2017}%
  \BibitemOpen
  \bibfield  {author} {\bibinfo {author} {\bibfnamefont {P.}~\bibnamefont
  {Bordia}}, \bibinfo {author} {\bibfnamefont {H.}~\bibnamefont
  {L{\"{u}}schen}}, \bibinfo {author} {\bibfnamefont {S.}~\bibnamefont
  {Scherg}}, \bibinfo {author} {\bibfnamefont {S.}~\bibnamefont
  {Gopalakrishnan}}, \bibinfo {author} {\bibfnamefont {M.}~\bibnamefont
  {Knap}}, \bibinfo {author} {\bibfnamefont {U.}~\bibnamefont {Schneider}}, \
  and\ \bibinfo {author} {\bibfnamefont {I.}~\bibnamefont {Bloch}},\
  }\href@noop {} {\bibfield  {journal} {\bibinfo  {journal} {Phys. Rev. X}\
  }\textbf {\bibinfo {volume} {7}},\ \bibinfo {pages} {041047} (\bibinfo {year}
  {2017})}\BibitemShut {NoStop}%
\bibitem [{\citenamefont {De~Tomasi}\ \emph {et~al.}(2017)\citenamefont
  {De~Tomasi}, \citenamefont {Bera}, \citenamefont {Bardarson},\ and\
  \citenamefont {Pollmann}}]{DeTomasi2017}%
  \BibitemOpen
  \bibfield  {author} {\bibinfo {author} {\bibfnamefont {G.}~\bibnamefont
  {De~Tomasi}}, \bibinfo {author} {\bibfnamefont {S.}~\bibnamefont {Bera}},
  \bibinfo {author} {\bibfnamefont {J.~H.}\ \bibnamefont {Bardarson}}, \ and\
  \bibinfo {author} {\bibfnamefont {F.}~\bibnamefont {Pollmann}},\ }\href
  {\doibase 10.1103/PhysRevLett.118.016804} {\bibfield  {journal} {\bibinfo
  {journal} {Phys. Rev. Lett.}\ }\textbf {\bibinfo {volume} {118}},\ \bibinfo
  {pages} {016804} (\bibinfo {year} {2017})}\BibitemShut {NoStop}%
\bibitem [{\citenamefont {Gray}\ \emph {et~al.}(2018)\citenamefont {Gray},
  \citenamefont {Bose},\ and\ \citenamefont {Bayat}}]{Gray2018}%
  \BibitemOpen
  \bibfield  {author} {\bibinfo {author} {\bibfnamefont {J.}~\bibnamefont
  {Gray}}, \bibinfo {author} {\bibfnamefont {S.}~\bibnamefont {Bose}}, \ and\
  \bibinfo {author} {\bibfnamefont {A.}~\bibnamefont {Bayat}},\ }\href@noop {}
  {\bibfield  {journal} {\bibinfo  {journal} {Phys. Rev. B}\ }\textbf {\bibinfo
  {volume} {97}},\ \bibinfo {pages} {201105} (\bibinfo {year}
  {2018})}\BibitemShut {NoStop}%
\bibitem [{\citenamefont {Bera}\ and\ \citenamefont
  {Lakshminarayan}(2016)}]{Bera2016}%
  \BibitemOpen
  \bibfield  {author} {\bibinfo {author} {\bibfnamefont {S.}~\bibnamefont
  {Bera}}\ and\ \bibinfo {author} {\bibfnamefont {A.}~\bibnamefont
  {Lakshminarayan}},\ }\href@noop {} {\bibfield  {journal} {\bibinfo  {journal}
  {Phys. Rev. B}\ }\textbf {\bibinfo {volume} {93}},\ \bibinfo {pages} {134204}
  (\bibinfo {year} {2016})}\BibitemShut {NoStop}%
\bibitem [{\citenamefont {Lim}\ and\ \citenamefont {Sheng}(2016)}]{Lim2016}%
  \BibitemOpen
  \bibfield  {author} {\bibinfo {author} {\bibfnamefont {S.~P.}\ \bibnamefont
  {Lim}}\ and\ \bibinfo {author} {\bibfnamefont {D.~N.}\ \bibnamefont
  {Sheng}},\ }\href@noop {} {\bibfield  {journal} {\bibinfo  {journal} {Phys.
  Rev. B}\ }\textbf {\bibinfo {volume} {94}},\ \bibinfo {pages} {045111}
  (\bibinfo {year} {2016})}\BibitemShut {NoStop}%
\bibitem [{\citenamefont {Luitz}\ \emph {et~al.}(2015)\citenamefont {Luitz},
  \citenamefont {Laflorencie},\ and\ \citenamefont {Alet}}]{Luitz2015}%
  \BibitemOpen
  \bibfield  {author} {\bibinfo {author} {\bibfnamefont {D.~J.}\ \bibnamefont
  {Luitz}}, \bibinfo {author} {\bibfnamefont {N.}~\bibnamefont {Laflorencie}},
  \ and\ \bibinfo {author} {\bibfnamefont {F.}~\bibnamefont {Alet}},\
  }\href@noop {} {\bibfield  {journal} {\bibinfo  {journal} {Phys. Rev. B}\
  }\textbf {\bibinfo {volume} {91}},\ \bibinfo {pages} {081103} (\bibinfo
  {year} {2015})}\BibitemShut {NoStop}%
\bibitem [{\citenamefont {Zhang}\ and\ \citenamefont {Yao}(2018)}]{Zhang2018}%
  \BibitemOpen
  \bibfield  {author} {\bibinfo {author} {\bibfnamefont {S.~X.}\ \bibnamefont
  {Zhang}}\ and\ \bibinfo {author} {\bibfnamefont {H.}~\bibnamefont {Yao}},\
  }\href@noop {} {\bibfield  {journal} {\bibinfo  {journal} {Phys. Rev. Lett.}\
  }\textbf {\bibinfo {volume} {121}},\ \bibinfo {pages} {206601} (\bibinfo
  {year} {2018})}\BibitemShut {NoStop}%
\bibitem [{\citenamefont {Vosk}\ \emph {et~al.}(2015)\citenamefont {Vosk},
  \citenamefont {Huse},\ and\ \citenamefont {Altman}}]{Vosk2015}%
  \BibitemOpen
  \bibfield  {author} {\bibinfo {author} {\bibfnamefont {R.}~\bibnamefont
  {Vosk}}, \bibinfo {author} {\bibfnamefont {D.~A.}\ \bibnamefont {Huse}}, \
  and\ \bibinfo {author} {\bibfnamefont {E.}~\bibnamefont {Altman}},\
  }\href@noop {} {\bibfield  {journal} {\bibinfo  {journal} {Phys. Rev. X}\
  }\textbf {\bibinfo {volume} {5}},\ \bibinfo {pages} {031032} (\bibinfo {year}
  {2015})}\BibitemShut {NoStop}%
\bibitem [{\citenamefont {Bera}\ \emph {et~al.}(2015)\citenamefont {Bera},
  \citenamefont {Schomerus}, \citenamefont {Heidrich-Meisner},\ and\
  \citenamefont {Bardarson}}]{Bera2015}%
  \BibitemOpen
  \bibfield  {author} {\bibinfo {author} {\bibfnamefont {S.}~\bibnamefont
  {Bera}}, \bibinfo {author} {\bibfnamefont {H.}~\bibnamefont {Schomerus}},
  \bibinfo {author} {\bibfnamefont {F.}~\bibnamefont {Heidrich-Meisner}}, \
  and\ \bibinfo {author} {\bibfnamefont {J.~H.}\ \bibnamefont {Bardarson}},\
  }\href@noop {} {\bibfield  {journal} {\bibinfo  {journal} {Phys. Rev. Lett.}\
  }\textbf {\bibinfo {volume} {115}},\ \bibinfo {pages} {046603} (\bibinfo
  {year} {2015})}\BibitemShut {NoStop}%
\bibitem [{\citenamefont {Ponte}\ \emph {et~al.}(2015)\citenamefont {Ponte},
  \citenamefont {Chandran}, \citenamefont {Papi{\'{c}}},\ and\ \citenamefont
  {Abanin}}]{Ponte2015}%
  \BibitemOpen
  \bibfield  {author} {\bibinfo {author} {\bibfnamefont {P.}~\bibnamefont
  {Ponte}}, \bibinfo {author} {\bibfnamefont {A.}~\bibnamefont {Chandran}},
  \bibinfo {author} {\bibfnamefont {Z.}~\bibnamefont {Papi{\'{c}}}}, \ and\
  \bibinfo {author} {\bibfnamefont {D.~A.}\ \bibnamefont {Abanin}},\ }\href
  {\doibase 10.1016/j.aop.2014.11.008} {\bibfield  {journal} {\bibinfo
  {journal} {Annals of Physics}\ }\textbf {\bibinfo {volume} {353}},\ \bibinfo
  {pages} {196} (\bibinfo {year} {2015})}\BibitemShut {NoStop}%
\bibitem [{\citenamefont {Bardarson}\ \emph {et~al.}(2012)\citenamefont
  {Bardarson}, \citenamefont {Pollmann},\ and\ \citenamefont
  {Moore}}]{Bardarson2012}%
  \BibitemOpen
  \bibfield  {author} {\bibinfo {author} {\bibfnamefont {J.~H.}\ \bibnamefont
  {Bardarson}}, \bibinfo {author} {\bibfnamefont {F.}~\bibnamefont {Pollmann}},
  \ and\ \bibinfo {author} {\bibfnamefont {J.~E.}\ \bibnamefont {Moore}},\
  }\href@noop {} {\bibfield  {journal} {\bibinfo  {journal} {Phys. Rev. Lett.}\
  }\textbf {\bibinfo {volume} {109}},\ \bibinfo {pages} {017202} (\bibinfo
  {year} {2012})}\BibitemShut {NoStop}%
\bibitem [{\citenamefont {{\v{Z}}nidari{\v{c}}}\ \emph
  {et~al.}(2008)\citenamefont {{\v{Z}}nidari{\v{c}}}, \citenamefont {Prosen},\
  and\ \citenamefont {Prelov{\v{s}}ek}}]{Znidaric2008}%
  \BibitemOpen
  \bibfield  {author} {\bibinfo {author} {\bibfnamefont {M.}~\bibnamefont
  {{\v{Z}}nidari{\v{c}}}}, \bibinfo {author} {\bibfnamefont {T.}~\bibnamefont
  {Prosen}}, \ and\ \bibinfo {author} {\bibfnamefont {P.}~\bibnamefont
  {Prelov{\v{s}}ek}},\ }\href@noop {} {\bibfield  {journal} {\bibinfo
  {journal} {Phys. Rev. B}\ }\textbf {\bibinfo {volume} {77}},\ \bibinfo
  {pages} {064426} (\bibinfo {year} {2008})}\BibitemShut {NoStop}%
\bibitem [{\citenamefont {Serbyn}\ \emph {et~al.}(2015)\citenamefont {Serbyn},
  \citenamefont {Papi\'{c}},\ and\ \citenamefont {Abanin}}]{Serbyn2015}%
  \BibitemOpen
  \bibfield  {author} {\bibinfo {author} {\bibfnamefont {M.}~\bibnamefont
  {Serbyn}}, \bibinfo {author} {\bibfnamefont {Z.}~\bibnamefont {Papi\'{c}}}, \
  and\ \bibinfo {author} {\bibfnamefont {D.~A.}\ \bibnamefont {Abanin}},\
  }\href {\doibase 10.1103/PhysRevX.5.041047} {\bibfield  {journal} {\bibinfo
  {journal} {Phys. Rev. X}\ }\textbf {\bibinfo {volume} {5}},\ \bibinfo {pages}
  {041047} (\bibinfo {year} {2015})}\BibitemShut {NoStop}%
\bibitem [{\citenamefont {Pietracaprina}\ \emph {et~al.}(2017)\citenamefont
  {Pietracaprina}, \citenamefont {Parisi}, \citenamefont {Mariano},
  \citenamefont {Pascazio},\ and\ \citenamefont
  {Scardicchio}}]{Pietracaprina2017}%
  \BibitemOpen
  \bibfield  {author} {\bibinfo {author} {\bibfnamefont {F.}~\bibnamefont
  {Pietracaprina}}, \bibinfo {author} {\bibfnamefont {G.}~\bibnamefont
  {Parisi}}, \bibinfo {author} {\bibfnamefont {A.}~\bibnamefont {Mariano}},
  \bibinfo {author} {\bibfnamefont {S.}~\bibnamefont {Pascazio}}, \ and\
  \bibinfo {author} {\bibfnamefont {A.}~\bibnamefont {Scardicchio}},\ }\href
  {\doibase 10.1088/1742-5468/aa9338} {\bibfield  {journal} {\bibinfo
  {journal} {Journal of Statistical Mechanics: Theory and Experiment}\ }\textbf
  {\bibinfo {volume} {2017}},\ \bibinfo {pages} {113102} (\bibinfo {year}
  {2017})}\BibitemShut {NoStop}%
\bibitem [{\citenamefont {Gray}\ \emph {et~al.}(2019)\citenamefont {Gray},
  \citenamefont {Bayat}, \citenamefont {Pal},\ and\ \citenamefont
  {Bose}}]{Gray2019}%
  \BibitemOpen
  \bibfield  {author} {\bibinfo {author} {\bibfnamefont {J.}~\bibnamefont
  {Gray}}, \bibinfo {author} {\bibfnamefont {A.}~\bibnamefont {Bayat}},
  \bibinfo {author} {\bibfnamefont {A.}~\bibnamefont {Pal}}, \ and\ \bibinfo
  {author} {\bibfnamefont {S.}~\bibnamefont {Bose}},\ }\href {\doibase
  10.48550/ARXIV.1908.02761} {\enquote {\bibinfo {title} {Scale invariant
  entanglement negativity at the many-body localization transition},}\ }
  (\bibinfo {year} {2019})\BibitemShut {NoStop}%
\bibitem [{\citenamefont {Luitz}\ \emph {et~al.}(2016)\citenamefont {Luitz},
  \citenamefont {Laflorencie},\ and\ \citenamefont {Alet}}]{Luitz2016}%
  \BibitemOpen
  \bibfield  {author} {\bibinfo {author} {\bibfnamefont {D.~J.}\ \bibnamefont
  {Luitz}}, \bibinfo {author} {\bibfnamefont {N.}~\bibnamefont {Laflorencie}},
  \ and\ \bibinfo {author} {\bibfnamefont {F.}~\bibnamefont {Alet}},\ }\href
  {\doibase 10.1103/PhysRevB.93.060201} {\bibfield  {journal} {\bibinfo
  {journal} {Phys. Rev. B}\ }\textbf {\bibinfo {volume} {93}},\ \bibinfo
  {pages} {060201} (\bibinfo {year} {2016})}\BibitemShut {NoStop}%
\bibitem [{\citenamefont {Smith}\ \emph {et~al.}(2016)\citenamefont {Smith},
  \citenamefont {Lee}, \citenamefont {Richerme}, \citenamefont {Neyenhuis},
  \citenamefont {Hess}, \citenamefont {Hauke}, \citenamefont {Heyl},
  \citenamefont {Huse},\ and\ \citenamefont {Monroe}}]{Smith2016}%
  \BibitemOpen
  \bibfield  {author} {\bibinfo {author} {\bibfnamefont {J.}~\bibnamefont
  {Smith}}, \bibinfo {author} {\bibfnamefont {A.}~\bibnamefont {Lee}}, \bibinfo
  {author} {\bibfnamefont {P.}~\bibnamefont {Richerme}}, \bibinfo {author}
  {\bibfnamefont {B.}~\bibnamefont {Neyenhuis}}, \bibinfo {author}
  {\bibfnamefont {P.~W.}\ \bibnamefont {Hess}}, \bibinfo {author}
  {\bibfnamefont {P.}~\bibnamefont {Hauke}}, \bibinfo {author} {\bibfnamefont
  {M.}~\bibnamefont {Heyl}}, \bibinfo {author} {\bibfnamefont {D.~A.}\
  \bibnamefont {Huse}}, \ and\ \bibinfo {author} {\bibfnamefont
  {C.}~\bibnamefont {Monroe}},\ }\href {\doibase 10.1038/nphys3783} {\bibfield
  {journal} {\bibinfo  {journal} {Nature Physics}\ }\textbf {\bibinfo {volume}
  {12}},\ \bibinfo {pages} {907} (\bibinfo {year} {2016})}\BibitemShut
  {NoStop}%
\bibitem [{\citenamefont {Sierant}\ \emph
  {et~al.}(2020{\natexlab{a}})\citenamefont {Sierant}, \citenamefont
  {Lewenstein},\ and\ \citenamefont {Zakrzewski}}]{Sierant2020pol}%
  \BibitemOpen
  \bibfield  {author} {\bibinfo {author} {\bibfnamefont {P.}~\bibnamefont
  {Sierant}}, \bibinfo {author} {\bibfnamefont {M.}~\bibnamefont {Lewenstein}},
  \ and\ \bibinfo {author} {\bibfnamefont {J.}~\bibnamefont {Zakrzewski}},\
  }\href {\doibase 10.1103/PhysRevLett.125.156601} {\bibfield  {journal}
  {\bibinfo  {journal} {Phys. Rev. Lett.}\ }\textbf {\bibinfo {volume} {125}},\
  \bibinfo {pages} {156601} (\bibinfo {year} {2020}{\natexlab{a}})}\BibitemShut
  {NoStop}%
\bibitem [{\citenamefont {Yao}\ \emph {et~al.}(2016)\citenamefont {Yao},
  \citenamefont {Laumann}, \citenamefont {Cirac}, \citenamefont {Lukin},\ and\
  \citenamefont {Moore}}]{Yao2016}%
  \BibitemOpen
  \bibfield  {author} {\bibinfo {author} {\bibfnamefont {N.~Y.}\ \bibnamefont
  {Yao}}, \bibinfo {author} {\bibfnamefont {C.~R.}\ \bibnamefont {Laumann}},
  \bibinfo {author} {\bibfnamefont {J.~I.}\ \bibnamefont {Cirac}}, \bibinfo
  {author} {\bibfnamefont {M.~D.}\ \bibnamefont {Lukin}}, \ and\ \bibinfo
  {author} {\bibfnamefont {J.~E.}\ \bibnamefont {Moore}},\ }\href@noop {}
  {\bibfield  {journal} {\bibinfo  {journal} {Phys. Rev. Lett.}\ }\textbf
  {\bibinfo {volume} {117}},\ \bibinfo {pages} {240601} (\bibinfo {year}
  {2016})}\BibitemShut {NoStop}%
\bibitem [{\citenamefont {Bera}\ \emph {et~al.}(2017)\citenamefont {Bera},
  \citenamefont {De~Tomasi}, \citenamefont {Weiner},\ and\ \citenamefont
  {Evers}}]{Bera2017}%
  \BibitemOpen
  \bibfield  {author} {\bibinfo {author} {\bibfnamefont {S.}~\bibnamefont
  {Bera}}, \bibinfo {author} {\bibfnamefont {G.}~\bibnamefont {De~Tomasi}},
  \bibinfo {author} {\bibfnamefont {F.}~\bibnamefont {Weiner}}, \ and\ \bibinfo
  {author} {\bibfnamefont {F.}~\bibnamefont {Evers}},\ }\href@noop {}
  {\bibfield  {journal} {\bibinfo  {journal} {Phys. Rev. Lett.}\ }\textbf
  {\bibinfo {volume} {118}},\ \bibinfo {pages} {196801} (\bibinfo {year}
  {2017})}\BibitemShut {NoStop}%
\bibitem [{\citenamefont {Doggen}\ \emph {et~al.}(2018)\citenamefont {Doggen},
  \citenamefont {Schindler}, \citenamefont {Tikhonov}, \citenamefont {Mirlin},
  \citenamefont {Neupert}, \citenamefont {Polyakov},\ and\ \citenamefont
  {Gornyi}}]{Doggen2018}%
  \BibitemOpen
  \bibfield  {author} {\bibinfo {author} {\bibfnamefont {E.~V.~H.}\
  \bibnamefont {Doggen}}, \bibinfo {author} {\bibfnamefont {F.}~\bibnamefont
  {Schindler}}, \bibinfo {author} {\bibfnamefont {K.~S.}\ \bibnamefont
  {Tikhonov}}, \bibinfo {author} {\bibfnamefont {A.~D.}\ \bibnamefont
  {Mirlin}}, \bibinfo {author} {\bibfnamefont {T.}~\bibnamefont {Neupert}},
  \bibinfo {author} {\bibfnamefont {D.~G.}\ \bibnamefont {Polyakov}}, \ and\
  \bibinfo {author} {\bibfnamefont {I.~V.}\ \bibnamefont {Gornyi}},\ }\href
  {\doibase 10.1103/PhysRevB.98.174202} {\bibfield  {journal} {\bibinfo
  {journal} {Phys. Rev. B}\ }\textbf {\bibinfo {volume} {98}},\ \bibinfo
  {pages} {174202} (\bibinfo {year} {2018})}\BibitemShut {NoStop}%
\bibitem [{\citenamefont {Chanda}\ \emph
  {et~al.}(2020{\natexlab{a}})\citenamefont {Chanda}, \citenamefont {Sierant},\
  and\ \citenamefont {Zakrzewski}}]{Chanda2020}%
  \BibitemOpen
  \bibfield  {author} {\bibinfo {author} {\bibfnamefont {T.}~\bibnamefont
  {Chanda}}, \bibinfo {author} {\bibfnamefont {P.}~\bibnamefont {Sierant}}, \
  and\ \bibinfo {author} {\bibfnamefont {J.}~\bibnamefont {Zakrzewski}},\
  }\href {\doibase 10.1103/PhysRevB.101.035148} {\bibfield  {journal} {\bibinfo
   {journal} {Phys. Rev. B}\ }\textbf {\bibinfo {volume} {101}},\ \bibinfo
  {pages} {035148} (\bibinfo {year} {2020}{\natexlab{a}})}\BibitemShut
  {NoStop}%
\bibitem [{\citenamefont {Chanda}\ \emph
  {et~al.}(2020{\natexlab{b}})\citenamefont {Chanda}, \citenamefont {Sierant},\
  and\ \citenamefont {Zakrzewski}}]{Chanda2020b}%
  \BibitemOpen
  \bibfield  {author} {\bibinfo {author} {\bibfnamefont {T.}~\bibnamefont
  {Chanda}}, \bibinfo {author} {\bibfnamefont {P.}~\bibnamefont {Sierant}}, \
  and\ \bibinfo {author} {\bibfnamefont {J.}~\bibnamefont {Zakrzewski}},\
  }\href {\doibase 10.1103/PhysRevResearch.2.032045} {\bibfield  {journal}
  {\bibinfo  {journal} {Phys. Rev. Research}\ }\textbf {\bibinfo {volume}
  {2}},\ \bibinfo {pages} {032045} (\bibinfo {year}
  {2020}{\natexlab{b}})}\BibitemShut {NoStop}%
\bibitem [{\citenamefont {Potter}\ \emph {et~al.}(2015)\citenamefont {Potter},
  \citenamefont {Vasseur},\ and\ \citenamefont {Parameswaran}}]{Potter2015}%
  \BibitemOpen
  \bibfield  {author} {\bibinfo {author} {\bibfnamefont {A.~C.}\ \bibnamefont
  {Potter}}, \bibinfo {author} {\bibfnamefont {R.}~\bibnamefont {Vasseur}}, \
  and\ \bibinfo {author} {\bibfnamefont {S.~A.}\ \bibnamefont {Parameswaran}},\
  }\href@noop {} {\bibfield  {journal} {\bibinfo  {journal} {Phys. Rev. X}\
  }\textbf {\bibinfo {volume} {5}},\ \bibinfo {pages} {031033} (\bibinfo {year}
  {2015})}\BibitemShut {NoStop}%
\bibitem [{\citenamefont {Chandran}\ \emph
  {et~al.}(2015{\natexlab{a}})\citenamefont {Chandran}, \citenamefont {Kim},
  \citenamefont {Vidal},\ and\ \citenamefont {Abanin}}]{Chandran2015liom}%
  \BibitemOpen
  \bibfield  {author} {\bibinfo {author} {\bibfnamefont {A.}~\bibnamefont
  {Chandran}}, \bibinfo {author} {\bibfnamefont {I.~H.}\ \bibnamefont {Kim}},
  \bibinfo {author} {\bibfnamefont {G.}~\bibnamefont {Vidal}}, \ and\ \bibinfo
  {author} {\bibfnamefont {D.~A.}\ \bibnamefont {Abanin}},\ }\href {\doibase
  10.1103/PhysRevB.91.085425} {\bibfield  {journal} {\bibinfo  {journal} {Phys.
  Rev. B}\ }\textbf {\bibinfo {volume} {91}},\ \bibinfo {pages} {085425}
  (\bibinfo {year} {2015}{\natexlab{a}})}\BibitemShut {NoStop}%
\bibitem [{\citenamefont {van Horssen}\ \emph {et~al.}(2015)\citenamefont {van
  Horssen}, \citenamefont {Levi},\ and\ \citenamefont
  {Garrahan}}]{VanHorssen2015}%
  \BibitemOpen
  \bibfield  {author} {\bibinfo {author} {\bibfnamefont {M.}~\bibnamefont {van
  Horssen}}, \bibinfo {author} {\bibfnamefont {E.}~\bibnamefont {Levi}}, \ and\
  \bibinfo {author} {\bibfnamefont {J.~P.}\ \bibnamefont {Garrahan}},\ }\href
  {\doibase 10.1103/PhysRevB.92.100305} {\bibfield  {journal} {\bibinfo
  {journal} {Phys. Rev. B}\ }\textbf {\bibinfo {volume} {92}},\ \bibinfo
  {pages} {100305} (\bibinfo {year} {2015})}\BibitemShut {NoStop}%
\bibitem [{\citenamefont {Brenes}\ \emph {et~al.}(2018)\citenamefont {Brenes},
  \citenamefont {Dalmonte}, \citenamefont {Heyl},\ and\ \citenamefont
  {Scardicchio}}]{Brenes2018}%
  \BibitemOpen
  \bibfield  {author} {\bibinfo {author} {\bibfnamefont {M.}~\bibnamefont
  {Brenes}}, \bibinfo {author} {\bibfnamefont {M.}~\bibnamefont {Dalmonte}},
  \bibinfo {author} {\bibfnamefont {M.}~\bibnamefont {Heyl}}, \ and\ \bibinfo
  {author} {\bibfnamefont {A.}~\bibnamefont {Scardicchio}},\ }\href {\doibase
  10.1103/PhysRevLett.120.030601} {\bibfield  {journal} {\bibinfo  {journal}
  {Phys. Rev. Lett.}\ }\textbf {\bibinfo {volume} {120}},\ \bibinfo {pages}
  {030601} (\bibinfo {year} {2018})}\BibitemShut {NoStop}%
\bibitem [{\citenamefont {Kuno}\ \emph {et~al.}(2020)\citenamefont {Kuno},
  \citenamefont {Orito},\ and\ \citenamefont {Ichinose}}]{Kuno2020}%
  \BibitemOpen
  \bibfield  {author} {\bibinfo {author} {\bibfnamefont {Y.}~\bibnamefont
  {Kuno}}, \bibinfo {author} {\bibfnamefont {T.}~\bibnamefont {Orito}}, \ and\
  \bibinfo {author} {\bibfnamefont {I.}~\bibnamefont {Ichinose}},\ }\href
  {\doibase 10.1088/1367-2630/ab6352} {\bibfield  {journal} {\bibinfo
  {journal} {New Journal of Physics}\ }\textbf {\bibinfo {volume} {22}},\
  \bibinfo {pages} {013032} (\bibinfo {year} {2020})}\BibitemShut {NoStop}%
\bibitem [{\citenamefont {L\"uschen}\ \emph {et~al.}(2017)\citenamefont
  {L\"uschen}, \citenamefont {Bordia}, \citenamefont {Scherg}, \citenamefont
  {Alet}, \citenamefont {Altman}, \citenamefont {Schneider},\ and\
  \citenamefont {Bloch}}]{Luschen2017}%
  \BibitemOpen
  \bibfield  {author} {\bibinfo {author} {\bibfnamefont {H.~P.}\ \bibnamefont
  {L\"uschen}}, \bibinfo {author} {\bibfnamefont {P.}~\bibnamefont {Bordia}},
  \bibinfo {author} {\bibfnamefont {S.}~\bibnamefont {Scherg}}, \bibinfo
  {author} {\bibfnamefont {F.}~\bibnamefont {Alet}}, \bibinfo {author}
  {\bibfnamefont {E.}~\bibnamefont {Altman}}, \bibinfo {author} {\bibfnamefont
  {U.}~\bibnamefont {Schneider}}, \ and\ \bibinfo {author} {\bibfnamefont
  {I.}~\bibnamefont {Bloch}},\ }\href {\doibase 10.1103/PhysRevLett.119.260401}
  {\bibfield  {journal} {\bibinfo  {journal} {Phys. Rev. Lett.}\ }\textbf
  {\bibinfo {volume} {119}},\ \bibinfo {pages} {260401} (\bibinfo {year}
  {2017})}\BibitemShut {NoStop}%
\bibitem [{\citenamefont {Chen}(2016)}]{Chen2016}%
  \BibitemOpen
  \bibfield  {author} {\bibinfo {author} {\bibfnamefont {Y.}~\bibnamefont
  {Chen}},\ }\href {\doibase 10.48550/ARXIV.1608.02765} {\enquote {\bibinfo
  {title} {Universal logarithmic scrambling in many body localization},}\ }
  (\bibinfo {year} {2016})\BibitemShut {NoStop}%
\bibitem [{\citenamefont {Lee}\ \emph {et~al.}(2019)\citenamefont {Lee},
  \citenamefont {Kim},\ and\ \citenamefont {Kim}}]{Lee2019}%
  \BibitemOpen
  \bibfield  {author} {\bibinfo {author} {\bibfnamefont {J.}~\bibnamefont
  {Lee}}, \bibinfo {author} {\bibfnamefont {D.}~\bibnamefont {Kim}}, \ and\
  \bibinfo {author} {\bibfnamefont {D.-H.}\ \bibnamefont {Kim}},\ }\href
  {\doibase 10.1103/PhysRevB.99.184202} {\bibfield  {journal} {\bibinfo
  {journal} {Phys. Rev. B}\ }\textbf {\bibinfo {volume} {99}},\ \bibinfo
  {pages} {184202} (\bibinfo {year} {2019})}\BibitemShut {NoStop}%
\bibitem [{\citenamefont {Xu}\ \emph {et~al.}(2019)\citenamefont {Xu},
  \citenamefont {Li}, \citenamefont {Hsu}, \citenamefont {Swingle},\ and\
  \citenamefont {Das~Sarma}}]{Xu2019}%
  \BibitemOpen
  \bibfield  {author} {\bibinfo {author} {\bibfnamefont {S.}~\bibnamefont
  {Xu}}, \bibinfo {author} {\bibfnamefont {X.}~\bibnamefont {Li}}, \bibinfo
  {author} {\bibfnamefont {Y.-T.}\ \bibnamefont {Hsu}}, \bibinfo {author}
  {\bibfnamefont {B.}~\bibnamefont {Swingle}}, \ and\ \bibinfo {author}
  {\bibfnamefont {S.}~\bibnamefont {Das~Sarma}},\ }\href {\doibase
  10.1103/PhysRevResearch.1.032039} {\bibfield  {journal} {\bibinfo  {journal}
  {Phys. Rev. Research}\ }\textbf {\bibinfo {volume} {1}},\ \bibinfo {pages}
  {032039} (\bibinfo {year} {2019})}\BibitemShut {NoStop}%
\bibitem [{\citenamefont {Chandran}\ \emph {et~al.}(2016)\citenamefont
  {Chandran}, \citenamefont {Pal}, \citenamefont {Laumann},\ and\ \citenamefont
  {Scardicchio}}]{Chandran2016}%
  \BibitemOpen
  \bibfield  {author} {\bibinfo {author} {\bibfnamefont {A.}~\bibnamefont
  {Chandran}}, \bibinfo {author} {\bibfnamefont {A.}~\bibnamefont {Pal}},
  \bibinfo {author} {\bibfnamefont {C.~R.}\ \bibnamefont {Laumann}}, \ and\
  \bibinfo {author} {\bibfnamefont {A.}~\bibnamefont {Scardicchio}},\ }\href
  {\doibase 10.1103/PhysRevB.94.144203} {\bibfield  {journal} {\bibinfo
  {journal} {Phys. Rev. B}\ }\textbf {\bibinfo {volume} {94}},\ \bibinfo
  {pages} {144203} (\bibinfo {year} {2016})}\BibitemShut {NoStop}%
\bibitem [{\citenamefont {Yu}\ \emph {et~al.}(2016)\citenamefont {Yu},
  \citenamefont {Luitz},\ and\ \citenamefont {Clark}}]{Yu2016}%
  \BibitemOpen
  \bibfield  {author} {\bibinfo {author} {\bibfnamefont {X.}~\bibnamefont
  {Yu}}, \bibinfo {author} {\bibfnamefont {D.~J.}\ \bibnamefont {Luitz}}, \
  and\ \bibinfo {author} {\bibfnamefont {B.~K.}\ \bibnamefont {Clark}},\ }\href
  {\doibase 10.1103/PhysRevB.94.184202} {\bibfield  {journal} {\bibinfo
  {journal} {Phys. Rev. B}\ }\textbf {\bibinfo {volume} {94}},\ \bibinfo
  {pages} {184202} (\bibinfo {year} {2016})}\BibitemShut {NoStop}%
\bibitem [{\citenamefont {Pekker}\ \emph {et~al.}(2017)\citenamefont {Pekker},
  \citenamefont {Clark}, \citenamefont {Oganesyan},\ and\ \citenamefont
  {Refael}}]{Pekker2017}%
  \BibitemOpen
  \bibfield  {author} {\bibinfo {author} {\bibfnamefont {D.}~\bibnamefont
  {Pekker}}, \bibinfo {author} {\bibfnamefont {B.~K.}\ \bibnamefont {Clark}},
  \bibinfo {author} {\bibfnamefont {V.}~\bibnamefont {Oganesyan}}, \ and\
  \bibinfo {author} {\bibfnamefont {G.}~\bibnamefont {Refael}},\ }\href
  {\doibase 10.1103/PhysRevLett.119.075701} {\bibfield  {journal} {\bibinfo
  {journal} {Phys. Rev. Lett.}\ }\textbf {\bibinfo {volume} {119}},\ \bibinfo
  {pages} {075701} (\bibinfo {year} {2017})}\BibitemShut {NoStop}%
\bibitem [{\citenamefont {Villalonga}\ and\ \citenamefont
  {Clark}(2020)}]{Villalonga2020}%
  \BibitemOpen
  \bibfield  {author} {\bibinfo {author} {\bibfnamefont {B.}~\bibnamefont
  {Villalonga}}\ and\ \bibinfo {author} {\bibfnamefont {B.~K.}\ \bibnamefont
  {Clark}},\ }\href {\doibase 10.48550/ARXIV.2005.13558} {\enquote {\bibinfo
  {title} {Eigenstates hybridize on all length scales at the many-body
  localization transition},}\ } (\bibinfo {year} {2020})\BibitemShut {NoStop}%
\bibitem [{\citenamefont {Holevo}(1973)}]{holevo1973}%
  \BibitemOpen
  \bibfield  {author} {\bibinfo {author} {\bibfnamefont {A.~S.}\ \bibnamefont
  {Holevo}},\ }\href@noop {} {\bibfield  {journal} {\bibinfo  {journal} {Probl.
  Peredachi Inf.}\ }\textbf {\bibinfo {volume} {9}},\ \bibinfo {pages} {3}
  (\bibinfo {year} {1973})}\BibitemShut {NoStop}%
\bibitem [{\citenamefont {Nielsen}\ and\ \citenamefont
  {Chuang}(2011)}]{nielsen2011}%
  \BibitemOpen
  \bibfield  {author} {\bibinfo {author} {\bibfnamefont {M.~A.}\ \bibnamefont
  {Nielsen}}\ and\ \bibinfo {author} {\bibfnamefont {I.~L.}\ \bibnamefont
  {Chuang}},\ }\href@noop {} {\emph {\bibinfo {title} {Quantum Computation and
  Quantum Information: 10th Anniversary Edition}}},\ \bibinfo {edition} {10th}\
  ed.\ (\bibinfo  {publisher} {Cambridge University Press},\ \bibinfo {address}
  {USA},\ \bibinfo {year} {2011})\BibitemShut {NoStop}%
\bibitem [{\citenamefont {Roga}\ \emph {et~al.}(2010)\citenamefont {Roga},
  \citenamefont {Fannes},\ and\ \citenamefont {Zyczkowski}}]{Roga2010}%
  \BibitemOpen
  \bibfield  {author} {\bibinfo {author} {\bibfnamefont {W.}~\bibnamefont
  {Roga}}, \bibinfo {author} {\bibfnamefont {M.}~\bibnamefont {Fannes}}, \ and\
  \bibinfo {author} {\bibfnamefont {K.}~\bibnamefont {Zyczkowski}},\
  }\href@noop {} {\bibfield  {journal} {\bibinfo  {journal} {Phys. Rev. Lett.}\
  }\textbf {\bibinfo {volume} {105}},\ \bibinfo {pages} {040505} (\bibinfo
  {year} {2010})}\BibitemShut {NoStop}%
\bibitem [{\citenamefont {Schumacher}\ and\ \citenamefont
  {Westmoreland}(1997)}]{Schumacher1997}%
  \BibitemOpen
  \bibfield  {author} {\bibinfo {author} {\bibfnamefont {B.}~\bibnamefont
  {Schumacher}}\ and\ \bibinfo {author} {\bibfnamefont {M.~D.}\ \bibnamefont
  {Westmoreland}},\ }\href {\doibase 10.1103/PhysRevA.56.131} {\bibfield
  {journal} {\bibinfo  {journal} {Phys. Rev. A}\ }\textbf {\bibinfo {volume}
  {56}},\ \bibinfo {pages} {131} (\bibinfo {year} {1997})}\BibitemShut
  {NoStop}%
\bibitem [{\citenamefont {Giovannetti}\ and\ \citenamefont
  {Fazio}(2005)}]{Giovannetti2005}%
  \BibitemOpen
  \bibfield  {author} {\bibinfo {author} {\bibfnamefont {V.}~\bibnamefont
  {Giovannetti}}\ and\ \bibinfo {author} {\bibfnamefont {R.}~\bibnamefont
  {Fazio}},\ }\href {\doibase 10.1103/PhysRevA.71.032314} {\bibfield  {journal}
  {\bibinfo  {journal} {Phys. Rev. A}\ }\textbf {\bibinfo {volume} {71}},\
  \bibinfo {pages} {032314} (\bibinfo {year} {2005})}\BibitemShut {NoStop}%
\bibitem [{\citenamefont {Macchiavello}\ \emph {et~al.}(2004)\citenamefont
  {Macchiavello}, \citenamefont {Palma},\ and\ \citenamefont
  {Virmani}}]{Macchiavello2004}%
  \BibitemOpen
  \bibfield  {author} {\bibinfo {author} {\bibfnamefont {C.}~\bibnamefont
  {Macchiavello}}, \bibinfo {author} {\bibfnamefont {G.~M.}\ \bibnamefont
  {Palma}}, \ and\ \bibinfo {author} {\bibfnamefont {S.}~\bibnamefont
  {Virmani}},\ }\href {\doibase 10.1103/PhysRevA.69.010303} {\bibfield
  {journal} {\bibinfo  {journal} {Phys. Rev. A}\ }\textbf {\bibinfo {volume}
  {69}},\ \bibinfo {pages} {010303} (\bibinfo {year} {2004})}\BibitemShut
  {NoStop}%
\bibitem [{\citenamefont {Lupo}\ \emph {et~al.}(2011)\citenamefont {Lupo},
  \citenamefont {Pirandola}, \citenamefont {Aniello},\ and\ \citenamefont
  {Mancini}}]{Lupo2011}%
  \BibitemOpen
  \bibfield  {author} {\bibinfo {author} {\bibfnamefont {C.}~\bibnamefont
  {Lupo}}, \bibinfo {author} {\bibfnamefont {S.}~\bibnamefont {Pirandola}},
  \bibinfo {author} {\bibfnamefont {P.}~\bibnamefont {Aniello}}, \ and\
  \bibinfo {author} {\bibfnamefont {S.}~\bibnamefont {Mancini}},\ }\href
  {\doibase 10.1088/0031-8949/2011/t143/014016} {\bibfield  {journal} {\bibinfo
   {journal} {Phys. Scr.}\ }\textbf {\bibinfo {volume} {T143}},\ \bibinfo
  {pages} {014016} (\bibinfo {year} {2011})}\BibitemShut {NoStop}%
\bibitem [{\citenamefont {Yang}\ \emph {et~al.}(2011)\citenamefont {Yang},
  \citenamefont {Bayat},\ and\ \citenamefont {Bose}}]{Yang2011}%
  \BibitemOpen
  \bibfield  {author} {\bibinfo {author} {\bibfnamefont {S.}~\bibnamefont
  {Yang}}, \bibinfo {author} {\bibfnamefont {A.}~\bibnamefont {Bayat}}, \ and\
  \bibinfo {author} {\bibfnamefont {S.}~\bibnamefont {Bose}},\ }\href {\doibase
  10.1103/PhysRevA.84.020302} {\bibfield  {journal} {\bibinfo  {journal} {Phys.
  Rev. A}\ }\textbf {\bibinfo {volume} {84}},\ \bibinfo {pages} {020302}
  (\bibinfo {year} {2011})}\BibitemShut {NoStop}%
\bibitem [{\citenamefont {Banchi}\ \emph {et~al.}(2017)\citenamefont {Banchi},
  \citenamefont {Fern\'andez-Rossier}, \citenamefont {Hirjibehedin},\ and\
  \citenamefont {Bose}}]{Banchi2017}%
  \BibitemOpen
  \bibfield  {author} {\bibinfo {author} {\bibfnamefont {L.}~\bibnamefont
  {Banchi}}, \bibinfo {author} {\bibfnamefont {J.}~\bibnamefont
  {Fern\'andez-Rossier}}, \bibinfo {author} {\bibfnamefont {C.~F.}\
  \bibnamefont {Hirjibehedin}}, \ and\ \bibinfo {author} {\bibfnamefont
  {S.}~\bibnamefont {Bose}},\ }\href {\doibase 10.1103/PhysRevLett.118.147203}
  {\bibfield  {journal} {\bibinfo  {journal} {Phys. Rev. Lett.}\ }\textbf
  {\bibinfo {volume} {118}},\ \bibinfo {pages} {147203} (\bibinfo {year}
  {2017})}\BibitemShut {NoStop}%
\bibitem [{\citenamefont {D'Arrigo}\ \emph {et~al.}(2015)\citenamefont
  {D'Arrigo}, \citenamefont {Benenti}, \citenamefont {Falci},\ and\
  \citenamefont {Macchiavello}}]{D'Arrigo2015}%
  \BibitemOpen
  \bibfield  {author} {\bibinfo {author} {\bibfnamefont {A.}~\bibnamefont
  {D'Arrigo}}, \bibinfo {author} {\bibfnamefont {G.}~\bibnamefont {Benenti}},
  \bibinfo {author} {\bibfnamefont {G.}~\bibnamefont {Falci}}, \ and\ \bibinfo
  {author} {\bibfnamefont {C.}~\bibnamefont {Macchiavello}},\ }\href {\doibase
  10.1103/PhysRevA.92.062342} {\bibfield  {journal} {\bibinfo  {journal} {Phys.
  Rev. A}\ }\textbf {\bibinfo {volume} {92}},\ \bibinfo {pages} {062342}
  (\bibinfo {year} {2015})}\BibitemShut {NoStop}%
\bibitem [{\citenamefont {Gray}(2018)}]{gray2018quimb}%
  \BibitemOpen
  \bibfield  {author} {\bibinfo {author} {\bibfnamefont {J.}~\bibnamefont
  {Gray}},\ }\href {\doibase 10.21105/joss.00819} {\bibfield  {journal}
  {\bibinfo  {journal} {Journal of Open Source Software}\ }\textbf {\bibinfo
  {volume} {3}},\ \bibinfo {pages} {819} (\bibinfo {year} {2018})}\BibitemShut
  {NoStop}%
\bibitem [{Note1()}]{Note1}%
  \BibitemOpen
  \bibinfo {note} {We take $t_\protect \text {Neel}=L$ for the entire
  letter.}\BibitemShut {Stop}%
\bibitem [{Note2()}]{Note2}%
  \BibitemOpen
  \bibinfo {note} {In this letter, we have used between $100$ and $1000$
  samples for each data point to get the disorder-averaged Holevo rate
  $\protect \bar {R}$.}\BibitemShut {Stop}%
\bibitem [{Note3()}]{Note3}%
  \BibitemOpen
  \bibinfo {note} {We take $T_0 = T_1/8$ for all cases.}\BibitemShut {Stop}%
\bibitem [{\citenamefont {Devakul}\ and\ \citenamefont
  {Singh}(2015)}]{Devakul2015}%
  \BibitemOpen
  \bibfield  {author} {\bibinfo {author} {\bibfnamefont {T.}~\bibnamefont
  {Devakul}}\ and\ \bibinfo {author} {\bibfnamefont {R.~R.~P.}\ \bibnamefont
  {Singh}},\ }\href {\doibase 10.1103/PhysRevLett.115.187201} {\bibfield
  {journal} {\bibinfo  {journal} {Phys. Rev. Lett.}\ }\textbf {\bibinfo
  {volume} {115}},\ \bibinfo {pages} {187201} (\bibinfo {year}
  {2015})}\BibitemShut {NoStop}%
\bibitem [{\citenamefont {{Harris}}(1974)}]{Harris1974}%
  \BibitemOpen
  \bibfield  {author} {\bibinfo {author} {\bibfnamefont {A.~B.}\ \bibnamefont
  {{Harris}}},\ }\href {\doibase 10.1088/0022-3719/7/9/009} {\bibfield
  {journal} {\bibinfo  {journal} {Journal of Physics C Solid State Physics}\
  }\textbf {\bibinfo {volume} {7}},\ \bibinfo {pages} {1671} (\bibinfo {year}
  {1974})}\BibitemShut {NoStop}%
\bibitem [{\citenamefont {Chayes}\ \emph {et~al.}(1986)\citenamefont {Chayes},
  \citenamefont {Chayes}, \citenamefont {Fisher},\ and\ \citenamefont
  {Spencer}}]{Chayes1986}%
  \BibitemOpen
  \bibfield  {author} {\bibinfo {author} {\bibfnamefont {J.~T.}\ \bibnamefont
  {Chayes}}, \bibinfo {author} {\bibfnamefont {L.}~\bibnamefont {Chayes}},
  \bibinfo {author} {\bibfnamefont {D.~S.}\ \bibnamefont {Fisher}}, \ and\
  \bibinfo {author} {\bibfnamefont {T.}~\bibnamefont {Spencer}},\ }\href
  {\doibase 10.1103/PhysRevLett.57.2999} {\bibfield  {journal} {\bibinfo
  {journal} {Phys. Rev. Lett.}\ }\textbf {\bibinfo {volume} {57}},\ \bibinfo
  {pages} {2999} (\bibinfo {year} {1986})}\BibitemShut {NoStop}%
\bibitem [{\citenamefont {Chandran}\ \emph
  {et~al.}(2015{\natexlab{b}})\citenamefont {Chandran}, \citenamefont
  {Laumann},\ and\ \citenamefont {Oganesyan}}]{Chandran2015clo}%
  \BibitemOpen
  \bibfield  {author} {\bibinfo {author} {\bibfnamefont {A.}~\bibnamefont
  {Chandran}}, \bibinfo {author} {\bibfnamefont {C.~R.}\ \bibnamefont
  {Laumann}}, \ and\ \bibinfo {author} {\bibfnamefont {V.}~\bibnamefont
  {Oganesyan}},\ }\href {\doibase 10.48550/ARXIV.1509.04285} {\enquote
  {\bibinfo {title} {Finite size scaling bounds on many-body localized phase
  transitions},}\ } (\bibinfo {year} {2015}{\natexlab{b}})\BibitemShut
  {NoStop}%
\bibitem [{\citenamefont {Panda}\ \emph {et~al.}(2020)\citenamefont {Panda},
  \citenamefont {Scardicchio}, \citenamefont {Schulz}, \citenamefont {Taylor},\
  and\ \citenamefont {{\v{Z}}nidari{\v{c}}}}]{Panda2020}%
  \BibitemOpen
  \bibfield  {author} {\bibinfo {author} {\bibfnamefont {R.~K.}\ \bibnamefont
  {Panda}}, \bibinfo {author} {\bibfnamefont {A.}~\bibnamefont {Scardicchio}},
  \bibinfo {author} {\bibfnamefont {M.}~\bibnamefont {Schulz}}, \bibinfo
  {author} {\bibfnamefont {S.~R.}\ \bibnamefont {Taylor}}, \ and\ \bibinfo
  {author} {\bibfnamefont {M.}~\bibnamefont {{\v{Z}}nidari{\v{c}}}},\ }\href
  {\doibase 10.1209/0295-5075/128/67003} {\bibfield  {journal} {\bibinfo
  {journal} {{EPL} (Europhysics Letters)}\ }\textbf {\bibinfo {volume} {128}},\
  \bibinfo {pages} {67003} (\bibinfo {year} {2020})}\BibitemShut {NoStop}%
\bibitem [{\citenamefont {Sierant}\ \emph
  {et~al.}(2020{\natexlab{b}})\citenamefont {Sierant}, \citenamefont
  {Delande},\ and\ \citenamefont {Zakrzewski}}]{Sierant2020}%
  \BibitemOpen
  \bibfield  {author} {\bibinfo {author} {\bibfnamefont {P.}~\bibnamefont
  {Sierant}}, \bibinfo {author} {\bibfnamefont {D.}~\bibnamefont {Delande}}, \
  and\ \bibinfo {author} {\bibfnamefont {J.}~\bibnamefont {Zakrzewski}},\
  }\href {\doibase 10.1103/PhysRevLett.124.186601} {\bibfield  {journal}
  {\bibinfo  {journal} {Phys. Rev. Lett.}\ }\textbf {\bibinfo {volume} {124}},\
  \bibinfo {pages} {186601} (\bibinfo {year} {2020}{\natexlab{b}})}\BibitemShut
  {NoStop}%
\bibitem [{\citenamefont {Dumitrescu}\ \emph {et~al.}(2019)\citenamefont
  {Dumitrescu}, \citenamefont {Goremykina}, \citenamefont {Parameswaran},
  \citenamefont {Serbyn},\ and\ \citenamefont {Vasseur}}]{Dumitrescu2019}%
  \BibitemOpen
  \bibfield  {author} {\bibinfo {author} {\bibfnamefont {P.~T.}\ \bibnamefont
  {Dumitrescu}}, \bibinfo {author} {\bibfnamefont {A.}~\bibnamefont
  {Goremykina}}, \bibinfo {author} {\bibfnamefont {S.~A.}\ \bibnamefont
  {Parameswaran}}, \bibinfo {author} {\bibfnamefont {M.}~\bibnamefont
  {Serbyn}}, \ and\ \bibinfo {author} {\bibfnamefont {R.}~\bibnamefont
  {Vasseur}},\ }\href {\doibase 10.1103/PhysRevB.99.094205} {\bibfield
  {journal} {\bibinfo  {journal} {Phys. Rev. B}\ }\textbf {\bibinfo {volume}
  {99}},\ \bibinfo {pages} {094205} (\bibinfo {year} {2019})}\BibitemShut
  {NoStop}%
\bibitem [{\citenamefont {Morningstar}\ \emph {et~al.}(2020)\citenamefont
  {Morningstar}, \citenamefont {Huse},\ and\ \citenamefont
  {Imbrie}}]{Morningstar2020}%
  \BibitemOpen
  \bibfield  {author} {\bibinfo {author} {\bibfnamefont {A.}~\bibnamefont
  {Morningstar}}, \bibinfo {author} {\bibfnamefont {D.~A.}\ \bibnamefont
  {Huse}}, \ and\ \bibinfo {author} {\bibfnamefont {J.~Z.}\ \bibnamefont
  {Imbrie}},\ }\href {\doibase 10.1103/PhysRevB.102.125134} {\bibfield
  {journal} {\bibinfo  {journal} {Phys. Rev. B}\ }\textbf {\bibinfo {volume}
  {102}},\ \bibinfo {pages} {125134} (\bibinfo {year} {2020})}\BibitemShut
  {NoStop}%
\bibitem [{\citenamefont {\v{S}untajs}\ \emph {et~al.}(2020)\citenamefont
  {\v{S}untajs}, \citenamefont {Bon\v{c}a}, \citenamefont {Prosen},\ and\
  \citenamefont {Vidmar}}]{Suntajs2020}%
  \BibitemOpen
  \bibfield  {author} {\bibinfo {author} {\bibfnamefont {J.}~\bibnamefont
  {\v{S}untajs}}, \bibinfo {author} {\bibfnamefont {J.}~\bibnamefont
  {Bon\v{c}a}}, \bibinfo {author} {\bibfnamefont {T.}~\bibnamefont {Prosen}}, \
  and\ \bibinfo {author} {\bibfnamefont {L.}~\bibnamefont {Vidmar}},\ }\href
  {\doibase 10.1103/PhysRevE.102.062144} {\bibfield  {journal} {\bibinfo
  {journal} {Phys. Rev. E}\ }\textbf {\bibinfo {volume} {102}},\ \bibinfo
  {pages} {062144} (\bibinfo {year} {2020})}\BibitemShut {NoStop}%
\bibitem [{\citenamefont {Laflorencie}\ \emph {et~al.}(2020)\citenamefont
  {Laflorencie}, \citenamefont {Lemari\'e},\ and\ \citenamefont
  {Mac\'e}}]{Laflorencie2020}%
  \BibitemOpen
  \bibfield  {author} {\bibinfo {author} {\bibfnamefont {N.}~\bibnamefont
  {Laflorencie}}, \bibinfo {author} {\bibfnamefont {G.}~\bibnamefont
  {Lemari\'e}}, \ and\ \bibinfo {author} {\bibfnamefont {N.}~\bibnamefont
  {Mac\'e}},\ }\href {\doibase 10.1103/PhysRevResearch.2.042033} {\bibfield
  {journal} {\bibinfo  {journal} {Phys. Rev. Research}\ }\textbf {\bibinfo
  {volume} {2}},\ \bibinfo {pages} {042033} (\bibinfo {year}
  {2020})}\BibitemShut {NoStop}%
\bibitem [{\citenamefont {Sels}\ and\ \citenamefont
  {Polkovnikov}(2021)}]{Sels2021}%
  \BibitemOpen
  \bibfield  {author} {\bibinfo {author} {\bibfnamefont {D.}~\bibnamefont
  {Sels}}\ and\ \bibinfo {author} {\bibfnamefont {A.}~\bibnamefont
  {Polkovnikov}},\ }\href {\doibase 10.1103/PhysRevE.104.054105} {\bibfield
  {journal} {\bibinfo  {journal} {Phys. Rev. E}\ }\textbf {\bibinfo {volume}
  {104}},\ \bibinfo {pages} {054105} (\bibinfo {year} {2021})}\BibitemShut
  {NoStop}%
\bibitem [{\citenamefont {Crowley}\ and\ \citenamefont
  {Chandran}(2020)}]{Crowley2020}%
  \BibitemOpen
  \bibfield  {author} {\bibinfo {author} {\bibfnamefont {P.~J.~D.}\
  \bibnamefont {Crowley}}\ and\ \bibinfo {author} {\bibfnamefont
  {A.}~\bibnamefont {Chandran}},\ }\href {\doibase 10.48550/ARXIV.2012.14393}
  {\enquote {\bibinfo {title} {A constructive theory of the numerically
  accessible many-body localized to thermal crossover},}\ } (\bibinfo {year}
  {2020})\BibitemShut {NoStop}%
\bibitem [{\citenamefont {Abanin}\ \emph {et~al.}(2021)\citenamefont {Abanin},
  \citenamefont {Bardarson}, \citenamefont {Tomasi}, \citenamefont
  {Gopalakrishnan}, \citenamefont {Khemani}, \citenamefont {Parameswaran},
  \citenamefont {Pollmann}, \citenamefont {Potter}, \citenamefont {Serbyn},\
  and\ \citenamefont {Vasseur}}]{Abanin2021}%
  \BibitemOpen
  \bibfield  {author} {\bibinfo {author} {\bibfnamefont {D.}~\bibnamefont
  {Abanin}}, \bibinfo {author} {\bibfnamefont {J.}~\bibnamefont {Bardarson}},
  \bibinfo {author} {\bibfnamefont {G.~D.}\ \bibnamefont {Tomasi}}, \bibinfo
  {author} {\bibfnamefont {S.}~\bibnamefont {Gopalakrishnan}}, \bibinfo
  {author} {\bibfnamefont {V.}~\bibnamefont {Khemani}}, \bibinfo {author}
  {\bibfnamefont {S.}~\bibnamefont {Parameswaran}}, \bibinfo {author}
  {\bibfnamefont {F.}~\bibnamefont {Pollmann}}, \bibinfo {author}
  {\bibfnamefont {A.}~\bibnamefont {Potter}}, \bibinfo {author} {\bibfnamefont
  {M.}~\bibnamefont {Serbyn}}, \ and\ \bibinfo {author} {\bibfnamefont
  {R.}~\bibnamefont {Vasseur}},\ }\href {\doibase 10.1016/j.aop.2021.168415}
  {\bibfield  {journal} {\bibinfo  {journal} {Annals of Physics}\ }\textbf
  {\bibinfo {volume} {427}},\ \bibinfo {pages} {168415} (\bibinfo {year}
  {2021})}\BibitemShut {NoStop}%
\bibitem [{Note4()}]{Note4}%
  \BibitemOpen
  \bibinfo {note} {This can be seen in the extreme for $l=L$ where $f(\cdot
  ,\cdot )$ is constant in all arguments.}\BibitemShut {Stop}%
\bibitem [{\citenamefont {Sorge}(2015)}]{sorge2015pyfssa}%
  \BibitemOpen
  \bibfield  {author} {\bibinfo {author} {\bibfnamefont {A.}~\bibnamefont
  {Sorge}},\ }\href {\doibase pyfssa 0.7.6. Zenodo.} {\enquote {\bibinfo
  {title} {pyfssa: a scientific python package for finite-size scaling analysis
  at phase transitions},}\ } (\bibinfo {year} {2015})\BibitemShut {NoStop}%
\bibitem [{\citenamefont {Melchert}(2009)}]{melchert2009autoscalepy}%
  \BibitemOpen
  \bibfield  {author} {\bibinfo {author} {\bibfnamefont {O.}~\bibnamefont
  {Melchert}},\ }\href {\doibase 10.48550/ARXIV.0910.5403} {\enquote {\bibinfo
  {title} {autoscale.py - a program for automatic finite-size scaling analyses:
  A user's guide},}\ } (\bibinfo {year} {2009})\BibitemShut {NoStop}%
\bibitem [{Note5()}]{Note5}%
  \BibitemOpen
  \bibinfo {note} {This latter suggestion is motivated by evidence that
  small-scale numerics violate the Harris criterion which insist that $\nu > 2$
  \cite {Harris1974, Chayes1986, Chandran2015clo}. For example, in Table~\ref
  {tab:critical}, the eigenstate environment types violate this bound the most,
  which may indicate that they are more susceptible to finite-size effects than
  the other two; driving down the value of $\nu $.}\BibitemShut {Stop}%
\end{thebibliography}
\end{document}